\RequirePackage{fix-cm}
\documentclass[smallextended]{svjour3}       
\smartqed  
\usepackage{graphicx}
\usepackage[dvipsnames]{xcolor}
\usepackage{array}
\usepackage{wrapfig}
\usepackage{multirow}
\usepackage{tabularx}
\usepackage{caption}
\usepackage{subcaption}
\usepackage{booktabs}
\usepackage{setspace}
\usepackage{tikz}
\usepackage{amssymb}
\usepackage{booktabs}
\usepackage{enumitem}
\usepackage{multirow}
\usepackage{longtable}
\usepackage{changepage}
\usepackage{hyperref}
\hypersetup{
    colorlinks=true,
    linkcolor=black,
    filecolor=black,      
    urlcolor=black,
    citecolor=black,
}
\usepackage{graphicx}

\usepackage{fancybox}
\usepackage{longtable}
\usepackage{tasks}
\usepackage{pdfpages}
\usepackage{fontawesome}
\journalname{Empirical Software Engineering}
\begin{document}

\title{Understanding the Influence of Motivation on Requirements Engineering-related Activities
}

\titlerunning{The Impact of Personality on RE}        

\author{Dulaji Hidellaarachchi       \and John Grundy \and Rashina Hoda \and Ingo Mueller 
}


\institute{D. Hidellaarachchi\at
              Dept. of Software Systems and Cybersecurity, Monash University, Melbourne, Australia \\
              \email{dulaji.hidellaarachchi@monash.edu}          
           \and
           J. Grundy \at
              Dept. of Software Systems and Cybersecurity, Monash University, Melbourne, Australia \\
              \email{john.grundy@monash.edu}  
              \and
              R.Hoda\at
              Dept. of Software Systems and Cybersecurity, Monash University, Melbourne, Australia \\
              \email{rashina.hoda@monash.edu} 
              \and
              I.Mueller \at
              Dept. of Software Systems and Cybersecurity, Monash University, Melbourne, Australia \\
              \email{ingo.mueller@monash.edu}   
}

\date{Received: date / Accepted: 18 July 2024}

\maketitle

\begin{abstract}
\emph{Context:} Requirements Engineering (RE)-related activities are critical in developing quality software and one of the most human-dependent processes in software engineering (SE). Hence, identifying the impact of diverse human-related aspects on RE is crucial in the SE context.
\emph{Objective:} Our study explores the impact of one of the most influential human aspects, \textbf{motivation} on RE, aiming to deepen understanding and provide practical guidance. \emph{Method:} By conducting semi-structured interviews with 21 RE-involved practitioners, we developed a theory using socio-technical grounded theory (STGT) that explains the contextual, causal, and intervening conditions influencing motivation in RE-related activities. \emph{Result:} We identified strategies to enhance motivating situations or mitigate demotivating ones, and the consequences resulting from applying these strategies. \emph{Conclusion:} Our findings offer actionable insights for software practitioners to manage the influence of motivation on RE and help researchers further investigate its role across various SE contexts in the future. 

\keywords{Motivation \and Requirements Engineering \and Software Engineering \and Human aspects \and Socio-Technical Grounded Theory}
\end{abstract}

\section{Introduction} \label{Intro}
Requirements Engineering (RE)-related activities, such as requirements elicitation, analysis, prioritization, validation and management are considered to be challenging yet important parts of the software development process as the success of the software projects mainly depends on developing what users want \cite{RN2731} \cite{sutcliffe2002user}. According to Capers Jones, in more than 75\% of all enterprises, RE is deficient and getting the right requirements is one of the most crucial yet difficult parts of a software project \cite{jones1991applied} \cite{tarawneh2011suggested}. The high rate of project failures emphasises the importance of understanding the causes behind the failures and the significance of RE in software engineering \cite{jones1991applied}. While there are a variety of reasons for software project success and/or failures,  \emph{``people"} are considered to be the major influence as SE is a human-intensive activity and humans play the central role in SE  \cite{murphy2010human}. Despite the fact that RE-related activities are highly human-centric, involving collaboration with a diverse range of people such as customers/clients, end-users, software developers, and requirements engineers, only a limited number of studies were found that specifically explore the impact of human aspects on RE-related activities \cite{john2005human} \cite{ali2019effective}.

\textbf{Motivation} has been identified as a prominent human aspect in completing software projects successfully and also as one of the most challenging to manage \cite{BEECHAM2008860} \cite{PROCACCINO2005194}. Motivation appears to be influenced by a diverse set of factors, ranging from context and development tasks to interpersonal dynamics of people. As a consequence, understanding the impact of motivation has become inherently complex yet crucial within the software engineering (SE) context \cite{FRANCA201479}. However, from
the systematic literature review (SLR) we conducted, we identified that \emph{motivation} is a human aspect that has received limited attention to date related to RE-related activities \cite{RN1600} and from the survey study we conducted with more than 100 software practitioners who are predominantly involved in RE identified motivation as the highest influential human aspect that impact RE \cite{10.1145/3546943}. 
This has been stated in numerous other related work where we identified that among diverse human aspects, motivation has been cited frequently as one of the causes of software project success and/or failure, yet there is limited focus on how it impacts on RE-related activities \cite{demarco2013peopleware}. 
Kolpondinos and Glinz \cite{kolpondinos2017tailoring} highlighted the importance of stakeholders' motivation on requirements elicitation in spite of having various requirements elicitation platforms such as Liquid RE \cite{7320433} or REfine \cite{snijders2015refine}. They further explained that, however, these platforms support the collaborative involvement of a number of stakeholders, which is not adequate in motivating them to actively contribute to requirements elicitation. This indicates that the stakeholders' motivation plays a crucial role in elicitation, and similarly, we wanted to identify how software practitioners' motivation impacts RE-related activities. 

To obtain a more comprehensive understanding of the impact of motivation on RE-related activities, we wanted to answer the research question \textbf{How does the motivation of software practitioners influence requirements engineering-related activities?}
To answer this research question, we conducted \textbf{21 in-depth interviews} with software practitioners who are predominantly involved in RE-related activities. By using socio-technical grounded theory method (STGT) and analysing the interviews in accordance with the basic and advanced stages in STGT \cite{RN1609}, we identified the impact of motivation on RE-related activities.  

The main contributions of this research are as follows:
\begin{itemize}
 \item we developed a theory  to describe the impact of motivation on RE-related activities; 
    \item discovered a set of factors that motivate \& demotivate software practitioners when involved in RE-related activities;
    \item discovered a set of practical strategies that can be used to reinforce motivating situations or to mitigate demotivating situations when involved in RE-related activities; and
    \item a set of recommendations for future research into the impact of motivation on RE and SE in general.
\end{itemize}

\section{Related Work} \label{Related Work}
\textbf{Motivation} is a human aspect that consists of various definitions from different disciplines. Although the term \emph{``motivation"} has more than 140 formal definitions, most of these definitions cover a common set of characteristics. For example, motivation is internal to individuals, varies due to their objectives and helps to determine human behaviour \cite{RN3000}. Among these,  our research study uses one of the most common definitions in the SE context: \textbf{Motivation} is  \emph{``the willingness to do a certain action and is conditioned by this action's ability to satisfy needs for the individual" -Robbins definition} \cite{RN3002}. Various motivational theories have been used in research studies that originated in the psychology research domain, such as, \emph{``Maslow's motivation theory"} \cite{4685653} \cite{sutcliffe2002user} \emph{``Motivation-Hygiene theory"} \cite{francca2010designing}, \emph{``Expectancy theory"} \cite{francca2010designing}, and \emph{``The MOCC model"} \cite{francca2014motivated}. In their study, Sharp et al. \cite{sharp2009models} discussed the use of various motivation models for software engineers' motivation. They found that the majority of these models focus on the management perspective, and none of the models considered the software engineers' characteristics or their personal context, making each model represent a partial view of motivation in the SE context. They proposed a motivation model (MOCC Model) referring to four key components: Motivators, Outcomes, Characteristics and Context in software engineering, which needs to be further evaluated as it is proposed based on the literature. França et al. \cite{francca2013motivation} conducted a qualitative case study on the motivation of software engineers. They identified that, although there are classical motivation theories such as  Maslow's motivation theory and motivation-hygiene theory, there is a need for motivation theories which are capable of identifying motivators/demotivators that are specific to the SE context. Further, they pointed out that task variety and technical challenges are the main drivers of motivation for software practitioners, making it difficult to understand existing motivation theories completely.

Sach et al. \cite{6092590} conducted semi-structured interviews with 13 software engineers and suggested that motivating factors in SE are evolving and the current understanding of motivation in SE might be out of date. They identified that ‘work that is useful', ‘producing good software’ and ‘solving problems’ emerged as the most cited motivating factors. A cross-case analysis of two software organizations was conducted by França et al. \cite{FRANCA201479}, focusing on building explanatory theories of motivation in different software organizations and integrating them to obtain a comprehensive understanding of the role of motivation in SE. They suggested conducting a deep investigation into the motivation of software engineers focusing on their role in the team as it would impact the performance of the team. In \cite{RN2713}, an empirical study was conducted to investigate how software testers can be better motivated. Semi-structured in-depth interviews were conducted with 36 practitioners in 12 software organizations in Norway. A set of motivating and demotivating factors influencing software testing personnel was identified and proposed that combining testing responsibilities with a variety of tasks engagement increases the satisfaction of testers which eventually increases their motivation. 

The SLR conducted by Beecham et al. \cite{BEECHAM2008860} on motivation in SE highlights that software engineers are likely to be motivated according to three related factors, namely; their ‘characteristics’ (their need for variety); internal ‘controls’ (their personality) and external ‘moderators’ (their career stage), but mentioned that these could differ with various roles in the software development process. Further, it emphasises that non of these prior work they included in the SLR has considered the diverse SE activities, which is the primary motivator for their tasks where they have mainly considered programming/development and management related tasks among these studies \cite{rasch1992factors} \cite{hertel2003motivation} \cite{richens1998hr} \cite{garden1988behavioural}. Therefore, they suggest conducting future studies considering diverse tasks in software projects to fulfill the need for comprehensive model of motivation in software engineering. Further, in \cite{RN2572} they elaborated that although there are a number of motivational theories that originated in the psychology domain that have been taken into consideration in the SE research (e.g Hackman and Oldham’s Job Characteristics theory \cite{hackman1976motivation}),  these classic motivation theories do not explicitly interpret software practitioners' motivation and emphasised the need for more empirical studies to develop a strong body of knowledge on motivation in the SE domain rather than limiting it to existing theories mostly derived from psychological domain. 

Khan and Akbar  \cite{RN2215} performed an SLR and an empirical investigation on motivation factors for the requirements change management process in global software development (GSD). They explored the motivators contributing to requirements change management by extracting 25 motivators and finally developed taxonomies of identified motivators such as accountability,  clear change management strategy, overseas site response, effective requirements change management leadership, etc. Most of these systematic and empirical studies to date have  focused on software engineering activities in general, or predominantly design, development, testing and GSD contexts. The studies that focused on RE have mainly been limited to the GSD domain. Considering all these studies and based on the findings of our own prior studies into the impact of human aspects on RE, we decided to conduct a set of semi-structured, in-depth interviews specifically targeting software practitioners involved in RE. We wanted to find out their key motivating/demotivating factors and the impacts of these on performing RE-related tasks.

\section{Research Methodology} \label{methodology}

 \begin{figure*}[htbp]
 \centering
  \includegraphics[width=\linewidth]{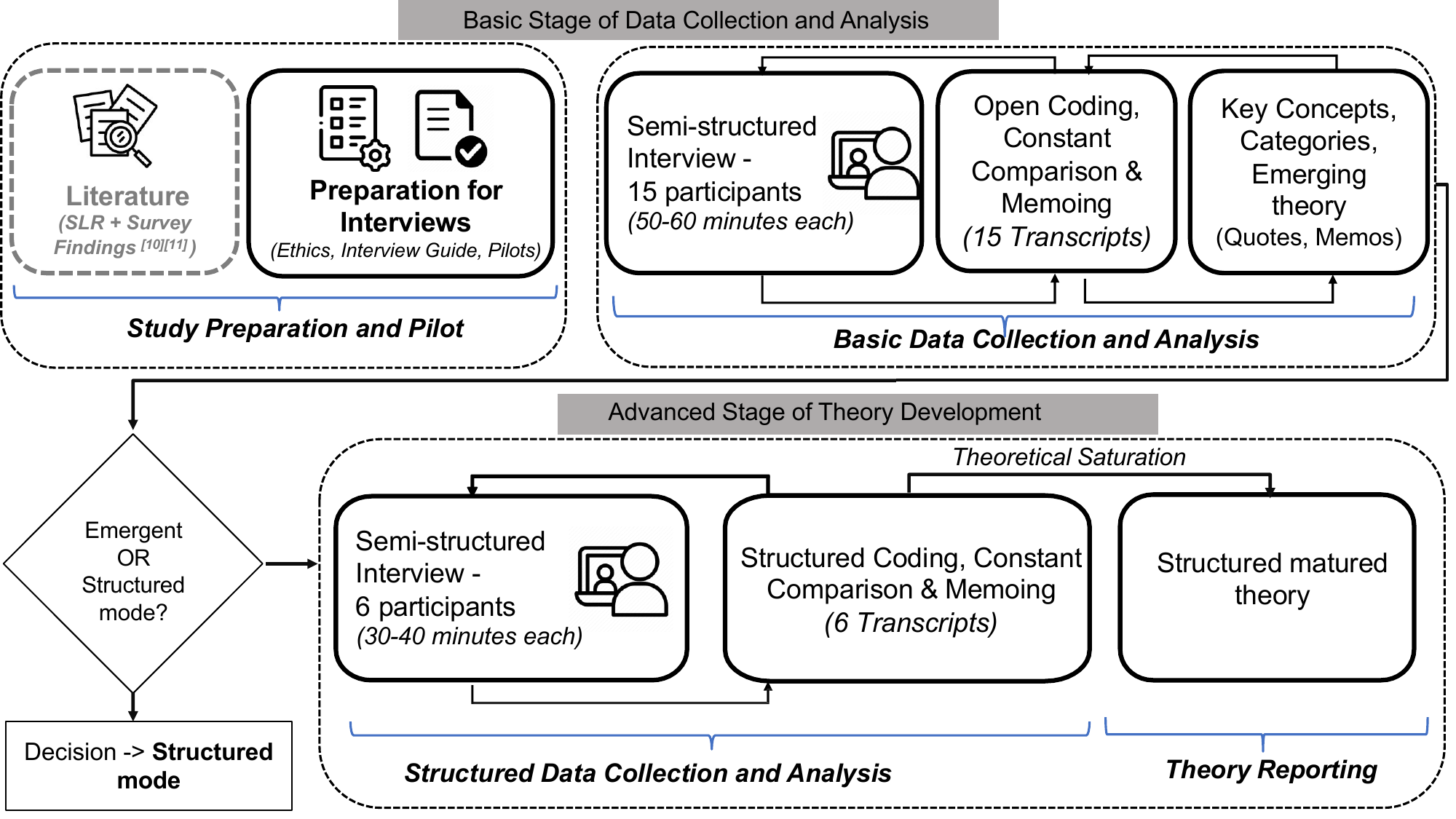}
  \caption{Overview of applying socio-technical grounded theory's basic and advanced stages}
  \label{Methodology figure}
\end{figure*}

We applied socio-technical grounded theory (STGT), which uses a primarily inductive and iterative approach to generate theory from data \cite{RN1609}. STGT is custom designed to delve deeply into phenomena that occur in socio-technical contexts such as the context of our study, particularly when seeking a comprehensive understanding through purposefully sampled qualitative data \cite{madampe2022role} \cite{10061282}. Hence, we selected STGT as our research parameters closely align with the four dimensions of the STGT research framework as follows;
\begin{itemize}
    \item Socio-technical (ST) phenomenon - Our investigation on the influence of motivation on RE is a socio-technical phenomenon as these diverse impacts are related to individual, technical or team related factors when doing RE. 
    \item ST domain and actors - RE-related activities in software projects are considered to be human-centered \cite{RN2731} \cite{sutcliffe2002user}, involving diverse actors such as software engineers, business analysts, project managers, customers, etc.
    \item ST researchers - Our team consists of experience and skills in conducting socio-technical research, domain knowledge, and strong qualitative research skills that enable us to interact and understand the context and what participants refer to. 
    \item ST data tools and techniques - We used a number of tools including Otter.ai for automated transcriptions, NVivo tool to assist with the coding and analysis of interviews, and Zoom recordings.
\end{itemize}

The methodology offers a systematic two-step approach, consisting of the \textit{basic stage} for data collection and analysis, followed by the \textit{advanced stage} focusing on theory development.  Figure \ref{Methodology figure} shows an overview of the STGT method applied in this research, including both the basic and advanced stages. Further details of applying STGT can be found in section \ref{Design} and \ref{analysis}. The sections below provide details of these steps.

\subsection{Basic Stage of Data Collection and Analysis} \label{Design}
\subsubsection{Study Preparation and Pilot}
Our study design was motivated by our previous research. Our SLR \cite{RN1600} identified that human aspects and their impact on RE-related activities have received limited attention. While \emph{motivation} has been studied in SE, its relation to RE activities is under-explored. To align our SLR findings with real-world experiences, we conducted a survey with 111 software practitioners focused on RE \cite{10.1145/3546943}. They identified \emph{motivation} as the most influential human aspect in RE activities.
These prior findings led us to focus on motivation impact on RE, and design our current study to elicit deeper insights into the impact of motivation when involved in RE-related activities from the software practitioners.
To obtain an in-depth understanding of the impact of motivation on RE-related activities,  we decided to conduct a full STGT study of real-world practice. We collected rich qualitative data through interviews targeting software practitioners involved in RE-related activities. Prior findings guided the design of our interview protocol.

We designed a semi-structured interview protocol with two sections. The first collected demographic information and details on participants' involvement in RE-related activities and project domains. 
The second section collected the participant's views on the factors that motivate them when involved in RE-related activities. In this section, we collected their views on the motivating/ demotivating factors and their impacts, their experiences of working with motivated \& demotivated teams and how they keep up the motivation of the team or work on the demotivated situations when involved in RE-related activities. The interview protocol is available in Appendix \ref{A}. 
Next, we conducted a pilot study with two software practitioners from our professional networks to validate the clarity and understandability of the questions; the time reported to complete the study and to elicit their suggestions on improving the interview study.  Both provided feedback, and we modified the questions to improve the clarity based on their suggestions and finalized the interview study. The data was collected from 21 practitioners who were mainly involved in RE-related activities. 

\subsubsection{Basic Data Collection and Analysis} \label{data collection}
The target population of our study was software practitioners involved in RE-related activities. We used purposive sampling to recruit those predominantly involved in RE to gather valuable insights into motivation. After obtaining the required ethics approval\footnote{Monash Ethics Review Manager (ERM) reference number: 29072}, we advertised our study on social media (LinkedIn, Twitter, Facebook) and within our professional networks. Participation was voluntary, and data collection and analysis were iterative. Initially, we collected data from 15 software practitioners worldwide, chosen based on their RE roles and experiences, including business analysts, software engineers, and other RE practitioners. Email conversations helped select suitable participants and obtain their consent.

\begin{figure}[htbp]
  \includegraphics[width=1.1\linewidth]{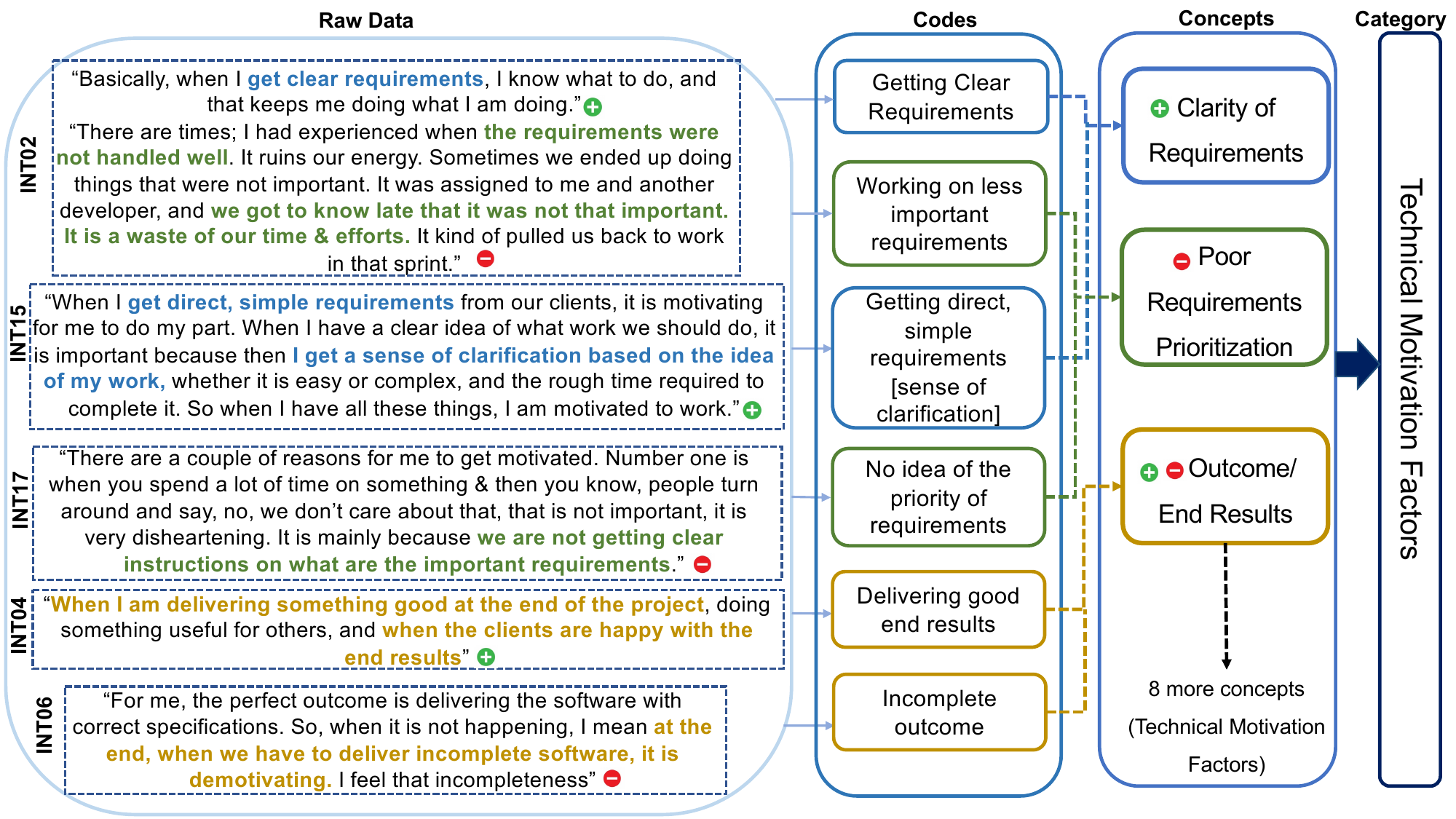}
  \caption{Example of STGT analysis, moving from raw data to codes to concepts to category, for the category ``Technical motivation factors" (The colour coding, ``\textbf{blue, green, yellow}" indicates the three different concepts identified analysing the raw data)}
  \label{STGT analysis}
\end{figure}

Interview participant recruitment was interleaved with analysis throughout the study and gathered qualitative data by conducting semi-structured interviews, with each interview lasting 50-60 minutes in the basic stage. We transcribed the interview sessions using Otter.ai with the consent of the participants, checked against the interview recordings to correct misinterpretations of the generated transcriptions (if any) and stored and analyzed the data using NViVO. We followed the open coding procedure to generate concepts and categories with constant comparison and memoing techniques. Figure \ref{STGT analysis} shows an example of STGT analysis and how it led to the emergence of the category \textit{"technical motivation factors"} with constant comparison and grouping of codes and concepts from raw data. We used open coding for the raw data (interview transcripts), and with constant comparison within the same and across different transcripts, we grouped similar codes to define concepts. Moreover, based on the explanations of the participants (raw data), these concepts were identified as positive (denoted by the icon \includegraphics[height=1em]{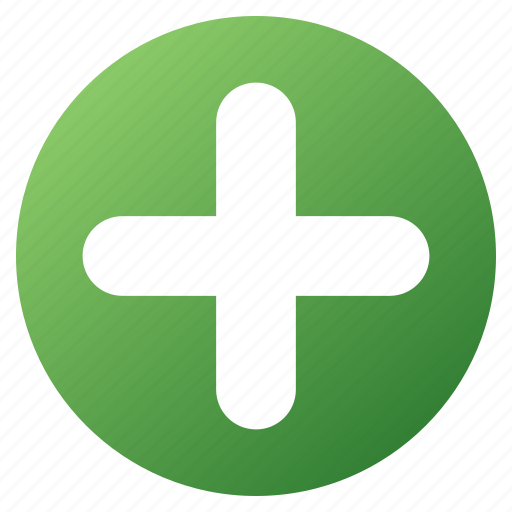})  and negative (denoted by the icon \includegraphics[height=1em]{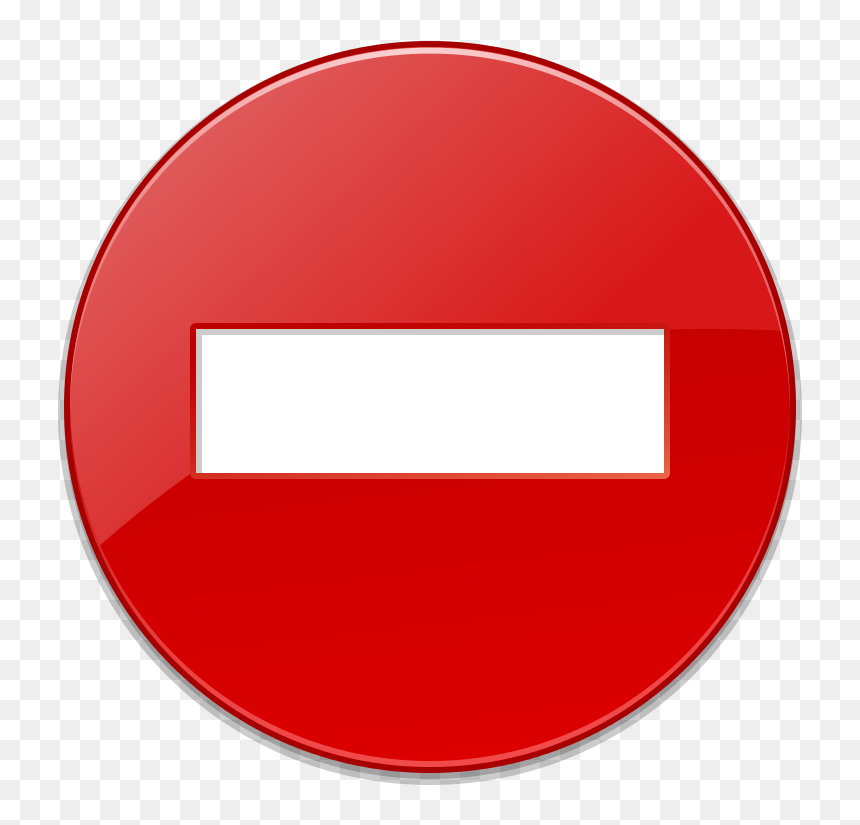}) where we were able to identify positively and/or negatively impacted motivation factors. 

According to the example provided in Figure \ref{STGT analysis}, these concepts were identified and grouped into the category of \textit{technical factors} that motivate practitioners when performing RE-related activities. Similarly, we identified a list of motivating/demotivating factors. During coding of each interview, we created ``memos". \textit{Basic memoing is the process of documenting the researcher’s thoughts, ideas, and reflections on emerging concepts and (sub)categories and evidence-based conjectures on possible links between them}'' \cite{RN1609}. We wrote \textbf{memos} to record key insights while following the open coding activities. Memos were focused around emerging categories and concepts. Below is an example of a memo that we recorded related to the "use of advanced technologies" as a motivation factor. We discuss insights generated from memoing in section \ref{section 5.2}.\\

\color{black}
\cornersize{.08} 
\ovalbox{\centering \begin{minipage}{30em}
\small
\par \textbf{Memo on ``Use of advanced tools \& techniques as a Motivation Factor''} of the 21 participants, 17 mentioned that \textit{use of various technologies} motivate them to perform RE-related activities. They have referred to it as \textit{``use of various tools" (INT01)}, \textit{``customized tools" (INT02)}, \textit{``use of advanced technologies" (INT06)}, or \textit{``tech-advanced applications for RE" (INT21)} act as a motivation factor for them to involve in RE-activities. These can be considered positive attributes that can influence practitioners' motivation. It is interesting to know how these attributes impact RE-related activities such as ``helpful in advancing requirements" (INT21), ``ability to embed a variety of functions" (INT09) and ``do customization according to requirements" (INT02), and ``enhance the system design in line with requirements" (INT09) mentioned by the practitioners. However, one practitioner (INT16) highlighted that technology should have a threshold; if it goes overboard,it may impact negatively as well. Various other aspects, such as practitioners' individual interests, the project domain, and practitioners' knowledge, play a vital role in considering the "use of various technologies" as a motivation factor to perform RE-related activities. Hence, it is important to understand how various technologies can be used for better RE with more context-specified future studies. 
\end{minipage}}\\
 
\subsection{Advanced Stage of Theory Development} \label{analysis}
Towards the end of our basic data collection and analysis (with 15 practitioners), key concepts, and categories that started to emerge which led us to step into the advanced theory development stage in STGT. The advanced stage provides two options for theory development, namely emergent or structured, both resulting in mature theory as an outcome. At this stage, we looked for possible pre-defined theoretical templates (if any) that would fit our emerging findings. We identified Strauss and Corbin's \textit{coding paradigm model} \cite{strauss1990basics} was best suited as a template for structuring and developing our emerging theory as it best fit our emerging findings from the basic stage. In other words, we used the coding paradigm model towards of the end of the basic stage of our STGT application to structure our emerging theory.

In this stage, we conducted structured data collection, where we focused on collecting data to progressively strengthen and saturate the concepts and categories we had identified in the basic stage as part of the emerging theory structure. It was mainly about saturating the existing categories and occasionally identifying new concepts that appear within the theoretical structure. For that, we collected data from six software practitioners working in the same software development team in an organization in Australia, making the total of 21 interview participants in our study. To recruit these six software practitioners, we contacted the team lead and invited the members who are mainly involved in RE to participate in our interview study. The interviews in this stage lasted around 30-40 minutes.

Across both stages, the core interview questions remained consistent to maintain the focus of our study. However, as the analysis of 15 interviews (basic data collection and analysis) revealed key concepts and categories, leading us to explore them further through follow-up questions in the advanced stage to strengthen our theory. This supported us to delve deeper into the nuanced aspects identified during the initial analysis, allowing for a more comprehensive understanding of the factors influencing motivation concerning RE activities and facilitating data saturation. For example, the last couple of interviews seem to validate our findings instead of adding any fundamentally new insights. As a result we could structure the concepts and relationships that we identified via STGT to the coding paradigm to visualise the theory. For example, all the concepts and categories that were identified as motivating/demotivating factors were grouped under the "intervening conditions" that facilitate or constraint our phenomenon.  

Altogether, we collected data from 21 software practitioners. All the interviews were conducted online via Zoom, and were audio recorded. As depicted in both stages (section \ref{Design} and \ref{analysis}), the data collection and analysis were iterative, where we conducted 15 interviews in the first stage and six in the advanced stage (Table \ref{TABLE: Interview participants' demographics}). 
The qualitative  data was analysed by the first author and shared among the rest of the authors to discuss each step followed and identify the best ways to present the findings. This process was applied to the entire data set, and this way, all the conditions, strategies, and consequences were identified (section \ref{Findings}). Over the course of the study, the team held several discussion rounds to finalize the identified concepts and categories based on the qualitative data and contributed to presenting the findings. The analysis gave rise to the theory of how motivation influences RE-related activities presented in Figure \ref{Theory}.

\section{Findings} \label{Findings}

\begin{figure}[htbp]
\includegraphics[angle=90,origin=c, width=1.25\linewidth]{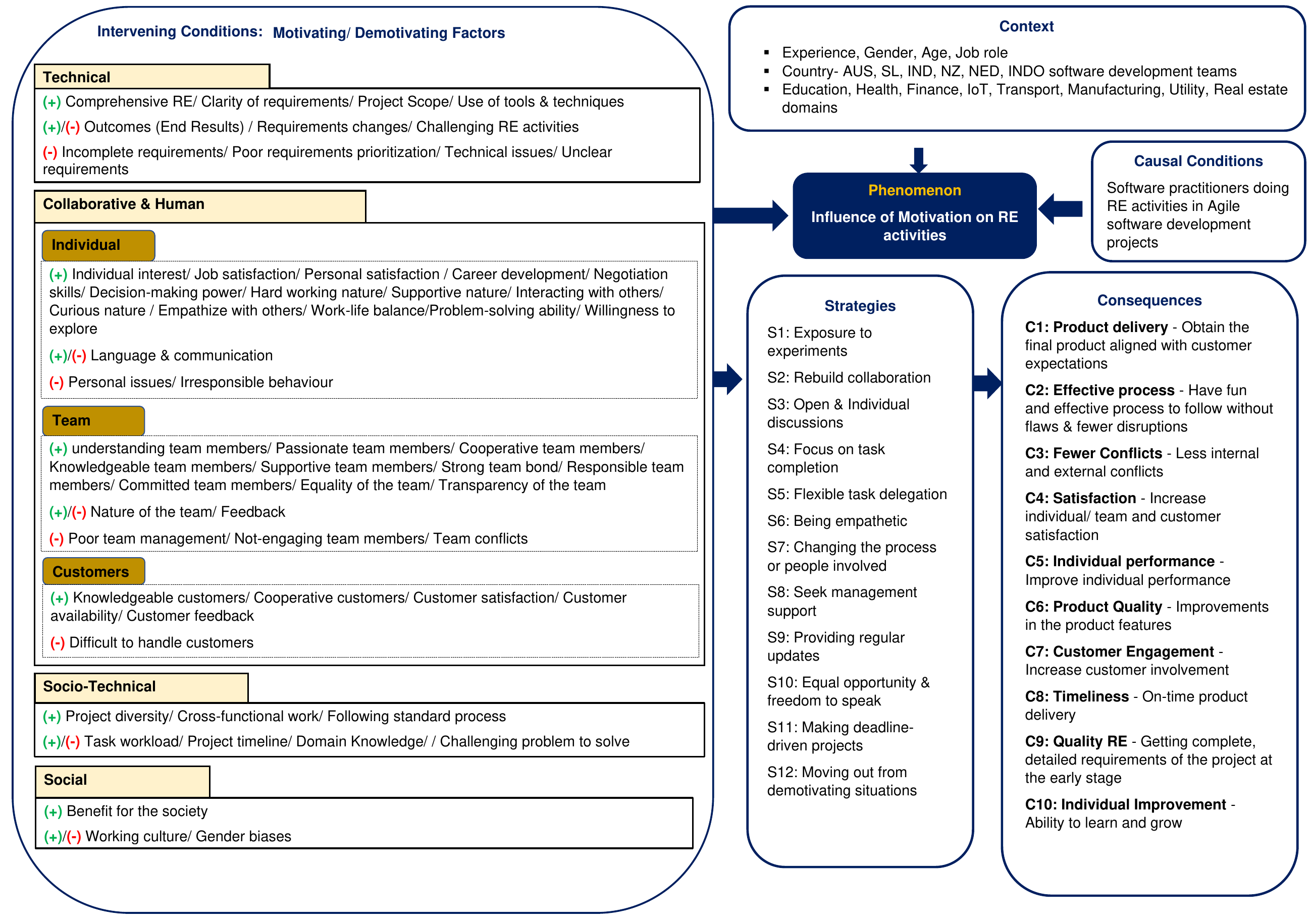}
   \caption{\centering The theory of the influence of motivation on RE-related activities, derived using Socio-Technical Grounded Theory (STGT) \cite{RN1609} and structured  using Coding Paradigm \cite{strauss1990basics}
   \footnotesize }
  \label{Theory}
\end{figure}

We present our theory of \textbf{the influence of motivation on RE-related activities}. 
Figure \ref{Theory} summarises it based on the analysis of the key findings from our interview study using STGT approach. We found how 
software practitioners' 
contexts along with
a set of causal and intervening conditions can give rise to or mediate the influence of motivation on RE-related activities. We also found a set of strategies applied by software practitioners to mitigate the demotivating situations when involved in RE, which result in a set of consequences. In the following sections, we describe each part of this theory -- Phenomenon (Influence of Motivation on RE-related activities); Context; Causal Conditions; Intervening Conditions; Strategies; and Consequences.

\subsection{The Phenomenon - Influence of Motivation on RE-related activities} \label{phenomenon}

In our study, the phenomenon we were investigating was ``the influence of motivation on RE-related activities". The findings highlighted the importance of having motivated software practitioners, how they influence RE-related activities and the overall software project. 
Several factors were identified as motivating/ demotivating factors that influence practitioners' motivation when involved in RE-related activities, varying based on their diverse contexts.
For example:\\
\emph{\faCommentsO ``Motivation is always related to the work I do, what type of work, the project I am doing, when I see it is beneficial for the people,  I get motivated. And I tend to contribute to the project more. I tend to carefully gather all the requirements, I don't want anything to be missed.
"} - Lead Business Analyst (INT11).\\
\emph{\faCommentsO ``For me, my motivation keeps changing based on the team I work with, the people I interact with, working on what I am good at, and of course my personal development. So these keep changing and it is difficult to find the perfect situation every time. 
"} - Business Analyst \& Project Manager (INT02).\\
Related to our phenomenon, we identified a set of factors that motivation is influenced by when involved in RE-related activities. 
These factors have a considerable impact on RE-related activities, and their influence results in various strategies and consequences used to handle motivation. 

 \begin{table*}[htbp]
\centering
\caption{\centering Demographics of the Interview Participants (*PID: Participant ID, *Ex. in SE: Experience in software industry) }
\label{TABLE: Interview participants' demographics}
\resizebox{\textwidth}{!}{%
\Large
\begin{tabular}{@{}lllllllll@{}}
\toprule
\multicolumn{1}{l}{\textbf{\begin{tabular}[c]{@{}l@{}}*PID\end{tabular} } }      & \textbf{Age}    & \textbf{Country}     & \textbf{\begin{tabular}[c]{@{}l@{}}*Ex.in \\SE\\(yrs)\end{tabular}}     &\textbf{\begin{tabular}[c]{@{}l@{}}Ex.in \\RE\\(yrs)\end{tabular}}  &\textbf{Job Role/Title}   & \textbf{Project Domains}    & \textbf{\begin{tabular}[c]{@{}l@{}}S/W Dev\\ Methods use\end{tabular}}                         \\ \midrule
\multicolumn{1}{l}{\textbf{\begin{tabular}[c]{@{}l@{}}\textit{Basic Stage}\end{tabular} } }      &  &    &    &     & &   &    &                         \\ \midrule

INT01 & 21-30 & Sri Lanka & 3.5 & 3.5 & Lead Business Analyst & \begin{tabular}[c]{@{}l@{}} IoT \&  Tele\\-communication \end{tabular} & Agile (Scrum) \\\\

INT02 & 21-30  & Sri Lanka & 6+ & 6+ & \begin{tabular}[c]{@{}l@{}}Business Analyst \& \\Project Manager \end{tabular}& \begin{tabular}[c]{@{}l@{}} Transport \& \\ Logistics\end{tabular} & Agile (scrum) \\\\

INT03 &  Above 50 & Indonesia & 24 & 18 & \begin{tabular}[c]{@{}l@{}}IT Development \\\& Re-engineering\\ Leader \end{tabular}& \begin{tabular}[c]{@{}l@{}}Manufacturing,\\ Web, ERP\end{tabular} & \begin{tabular}[c]{@{}l@{}}Agile \& \\Waterfall \end{tabular}\\\\

INT04 & 31-40  & Sri Lanka & 4 & 4 & Business Analyst & Finance, IT & Agile (scrum) \\\\

INT05 & 31-40 & Netherlands & 11 & 5+ & \begin{tabular}[c]{@{}l@{}}Software Engineer \& \\Application Consultant \end{tabular}& Telecommunication & Agile (scrum)\\\\

INT06 & 31-40  & New Zealand & 10 & 10 & Technical Team Lead & \begin{tabular}[c]{@{}l@{}}Finance, Insurance, \\Health \end{tabular} & \begin{tabular}[c]{@{}l@{}}Agile \\(2 weeks sprints)\end{tabular}\\\\

INT07 & 21-30  & Nepal & 5 & 2 & \begin{tabular}[c]{@{}l@{}}Software Engineer\\ (IoT, DevOps) \end{tabular}& \begin{tabular}[c]{@{}l@{}} Transport \& \\ Logistics\end{tabular} & Agile (scrum)\\\\

INT08 & 21-30  & Australia & 5 & 2 & Software Engineer & Real estate & Agile (scrum) \\\\

INT09 & Above 50  & Australia & 26 & 26 & \begin{tabular}[c]{@{}l@{}}Lead Human-centered \\Designer/ Information \\Architecture\end{tabular} & \begin{tabular}[c]{@{}l@{}} Transport \& \\ Logistics\end{tabular}& \begin{tabular}[c]{@{}l@{}}Agile \& \\Waterfall \end{tabular} \\\\

INT10 & 21-30  & India & 9 & 5 & Business Analyst & Utility services & Agile (scrum) \\
INT11 & Above 50  & New Zealand & 20+ & 20+ & Lead Business Analyst & Health & Agile (scrum)\\\\

INT12 & Above 50  & Australia & 26+ & 26+ &\begin{tabular}[c]{@{}l@{}}  Senior Project\\ Manager \end{tabular}& Health, Insurance & Agile (scrum)\\\\

INT13 & 21-30  & Sri Lanka & 5 & 5 & \begin{tabular}[c]{@{}l@{}}Lead Business Analyst, \\Scrum Master \end{tabular} & Health & Agile (scrum)\\\\

INT14 & 31-40  & India & 9 & 6 & Lead Business Analyst &  Health & Agile (scrum)\\\\

INT15 & 31-40  & India & 6+ & 5 & \begin{tabular}[c]{@{}l@{}}Software Engineer \\(SSE)\end{tabular} &  \begin{tabular}[c]{@{}l@{}} IoT \& Tele\\-communication \end{tabular} & Agile (scrum) \\ \midrule

\multicolumn{1}{l}{\textbf{\begin{tabular}[c]{@{}l@{}}\textit{Advanced stage}\end{tabular} } }      &  &    &    &     & &   &    &                         \\ \midrule

INT16 & 21-30 & Australia & 3 & 3 & Business Analyst &  Education & Agile (scrum) \\\\

INT17 & Above 50  & Australia & 25+ & 15 &  \begin{tabular}[c]{@{}l@{}}Lead Business Analyst, \\ Product Owner \end{tabular} &  Education & Agile (scrum) \\\\

INT18 & 31-40  & Australia & 10+ & 6+ &  \begin{tabular}[c]{@{}l@{}} Software Engineer \end{tabular} &  Education & Agile (scrum) \\\\

INT19 & 31-40  & Australia & 5 & 3 &  \begin{tabular}[c]{@{}l@{}} Business Analyst \end{tabular} &  Education & Agile (scrum) \\\\

INT20 & 41-50  & Australia & 25 & 20+ &  \begin{tabular}[c]{@{}l@{}} Senior Business Analyst \end{tabular} &  Education & Agile (scrum) \\\\

INT21 & 31-40  & Australia & 10+ & 7+ &  \begin{tabular}[c]{@{}l@{}} Project Manager/\\ Iteration Manager \end{tabular} &  Education & Agile (scrum) \\
\bottomrule
\end{tabular}%
}

\end{table*}

\subsection{The Context} \label{section 4.2}

The context can be described as "the specific set of characteristics in which the phenomenon is embedded," characterizing the conditions where strategies are applied to handle or react to a phenomenon \cite{strauss1990basics}. In our study, we define our context as a specific set of conditions  where the strategies can be made and applied related to the phenomenon which includes practitioners' team, project information (domains), the county of residence, their experience and job roles along with their age and gender. We consider these factors as context because they provide the backdrop within which the phenomena occur. Understanding these contextual factors helps contextualize the findings and provides insights into how motivation might be perceived or practiced based on geographical or industry-specific contexts. These elements influence the broader environment in which RE-related activities occur, impacting how motivation manifests in these settings.

Our 21 interview participants have software industry experience ranging from 3 to 26+ years, with most fully involved in RE-related activities. Detailed context information is provided in Table \ref{TABLE: Interview participants' demographics}. Of the participants, 12 are male, and the majority are in the 21-30 or 31-40 age categories. In the first iteration, 15 participants came from Sri Lanka (4), Australia (3), India (3), New Zealand (2), and one each from the Netherlands, Indonesia, and Nepal, working in diverse software development teams and project domains. Most worked in health (5), followed by transport \& logistics, IoT \& telecommunication (3 each), finance (2), insurance (2), and one each in manufacturing, real estate, utility services, and IT. In the second iteration, all 6 participants were from one software development team in Australia, working on a project in the education domain. Among the 21 participants, six held multiple roles, such as business analyst \& project manager (INT02), software engineer \& application consultant (INT05), and lead business analyst \& scrum master (INT13). The majority were Lead Business Analysts, Business Analysts, and Software Engineers (23.80\% each; 5 each), with others in roles like project manager, technical team lead, senior business analyst, consultant, lead human-centered designer, scrum master, and product owner. Understanding these contextual factors helped contextualize the findings and provided insights into how motivation is perceived or practiced in diverse RE settings.

\subsection{Causal Conditions - Practitioners doing RE-related activities in Agile Software Development Projects} \label{section 4.3}

Causal conditions specify "the phenomenon with respect to incidents or occurrences that result in the appearance or development of a phenomenon" \cite{strauss1990basics}. In our study, these conditions center on participants' involvement in RE-related activities within Agile software development projects. The rationale behind this designation draws upon the inherent characteristics of Agile methodologies, emphasizing constant communication, adaptability, and iterative development cycles. These unique features can potentially shape the motivation levels of practitioners engaged in RE-related activities. Our study revealed that all participants primarily employed Agile methodologies, mainly Scrum, while only two occasionally used traditional methods like waterfall. When discussing their motivation, they only provided examples from Agile projects, highlighting that the development method directly influences their motivation. Agile's iterative nature, regular feedback, and continuous improvement may enhance or diminish motivation in RE-related activities.

To understand this, we asked about the software development methodology used in projects and their involvement in RE-related activities. All participants were frequently involved in requirements elicitation, prioritization, and management. 90.47\% (n=19) were involved in requirements specification, analysis, and validation. Five participants mentioned activities like documenting software requirements and leading requirements analysis and specification work. Practitioners noted that engaging in RE within an Agile environment directly influences their motivation due to clear project objectives and faster feedback. Thus, we identified Agile methodology as the causal condition of our study. 
\\
\faCommentsO \emph{``...it's mainly iterative and incremental for that one project. Actually, all the other projects we use Scrum and I think it is huge different, I mean the method itself [is] a motivating thing for me, because then I can do my job well" - INT11 (Lead business analyst)}.

\subsection{Intervening Conditions - Motivating/ Demotivating Factors that impact RE-related activities} \label{section 4.4}

Intervening conditions are defined as ``the broad and general conditions that influence the strategies taken within a specific context" \cite{strauss1990basics}. In our study, these are factors that facilitate or constrain our phenomenon, specifically motivating or demotivating factors impacting RE-related activities and influencing the strategies used to manage motivation.
We found that software practitioners are motivated by various factors, which can have positive, negative, or mixed impacts. Figure \ref{Theory} highlights these key intervening conditions as motivating and demotivating factors on RE-related tasks. Most factors have a positive influence (denoted by \textbf{(+)}), some have both positive and negative influences (\textbf{(+/-)}), and a few have a negative influence (\textbf{(-)}). These factors and their influences are detailed in the following sub-sections, supported by original interview quotes.

\subsubsection{Technical Factors} \label{section 4.4.1}
In our study, we define \textbf{technical factors} as those related to the outcomes of RE-related activities. From responses to the interview question, \emph{``In your opinion, what are the aspects that motivate you when involved in RE-related activities?"} 
we identified a set of technical factors as a set of motivating/demotivating factors. 

Having \textbf{comprehensive requirements} is identified as one of the most positively influenced motivation factors, where nearly all of our participants (19 out of 21) mentioned that having detailed information about all the requirements (detailed information for the overall project) is vital as it motivates them to involve in the software development process and make the later tasks easier. Further, they elaborated that comprehensive requirements are important for the development team to understand the problem and design the most suitable solution while identifying priorities and dependencies of the requirements. Also, they highlighted that comprehensive RE led them to not miss any requirements, resulting in getting complete requirements in their projects that may affect the later part of the project. It is also helpful in preparing complete requirements specification documents, and as a result, sudden changes in the development team will not affect the project as team members do not have to rely on others. 
\par \emph{\faCommentsO ``For me, I always look for delivering the project with correct specifications. So, to do that I need to focus on getting all the requirements to make them workable functions, And, it should be stated from the beginning of the project, like getting all the requirements from the client, documenting them completely kind of motivating me. 
" -}  Senior Software Engineer (INT15)

\par \textbf{Clarity of requirements} is another positively influencing motivation factor, explained as clear, well-thought-out requirements that motivate practitioners to perform RE-related activities, resulting in better requirements analysis, improvement of the features, and massively reducing the pain, efforts or disruption in later phases of the software development cycle. 18 out of 21 participants mentioned that having clear requirements helps them to understand the exact requirements and whether they are on the right track to get the desired outcome. This is mainly referred to explicitness and precision of individual requirements emphasising the need for clear, unambiguous details for each specific task.
\par \emph{\faCommentsO ``Basically, when I get clear requirements, I know what to do, and that keeps me doing what I am doing. When I understand the requirements clearly, what the clients actually want, it is very easy for me to communicate it with the development team, and I know I am  doing it right" - } Business Analyst \& Project Manager (INT02)
\par \emph{\faCommentsO ``The most challenging part when you are doing requirements engineering is actually to do the analysis. So, to do it, you have to understand what is the exact requirement as an overview and in detail. If I have that exact details, analysis is not challenging for me, and I can do better work" -} Software Engineer \& Application Consultant (INT05)

\par The opposite of this condition, \textbf{unclear requirements}, is highlighted as a  key demotivating factor by the participants as it negatively influences RE-related activities. Vague and unclear requirements lead to a potential risk of developing less important features, resulting in a waste of resources in the project.
Almost half -- 10 out of 21 participants -- mentioned that they get confused or distracted by unclear requirements and are unable to make progress until they get clarification, impacting the project outcome, timeline and budget.  
\par \emph{\faCommentsO ``Sometimes, there are instances like the clients need something, but they don't know why they need it. -- in one of the projects, the client that I was talking to, asked for more dashboards, but he has no idea of what types of critical data they can provide and what do they need to see. So we had to arrange more discussion rounds to get it cleared. 
" -} Lead Business Analyst (INT01)

\par \textbf{Project scope} is also identified as a positively influenced motivation factor by twelve out of 21 participants). They highlighted that working on large-scale projects gives them the feeling of doing something big, with big contributions, and it motivates them to perform well with their allocated tasks. They believe that working on big-scaled projects provides lots of exposure to new techniques to elicit requirements and learn new technologies. Further, our participants shared that it brings energy to the team as everyone in the team gets to play various roles and collaborate with many people, making the whole development process exciting for them. 
\par \emph{\faCommentsO ``Within one project, we got to redesign seven different websites completely. It is motivating for me because it is a big task. For each one, the requirements are different and we had discussions with everyone who was handling those sites before we come up with a new design.  It was challenging as well as interesting because we used different approaches to collect requirements such as workshops, interviews, brainstorming sessions.." -}  Lead Human-centered Designer (INT09)

\par \textbf{Use of advanced tools and techniques} helps to make practitioners more motivated, resulting in positive influences on RE-related activities. 17 out of 21 participants explained that they get motivated when they can use advanced technologies \& various techniques as it is beneficial  for them in understanding requirements better, enhancing solutions/designs, advancing features according to requirements and managing requirement changes in later phases of the project. 
\par \emph{\faCommentsO ``I have been using different techniques and tools such as illustration models, some diagramming techniques to understand and analyse requirements. By using them, I get to find the most important artefacts, when it comes to the requirements we need. And I use it to come up with better solutions. It is kind of motivating for me" -} Lead Business Analyst (INT01)
\par \emph{\faCommentsO ``When your team is good at using advanced technologies, it is helpful to embed things into the project, like doing some customization without affecting to the whole project. It is a plus point when we have to handle changes. Because for some systems that we develop, it is very difficult to do changes after one point, because of the dependencies and all. So, that's where we need the technology, have to be updated with current techniques" -} Technical Team Lead (INT06)

\par \textbf{Outcomes (End results)} is a motivation factor that has both positive and negative influence according to nearly all of our participants (20 out of 21).  The usefulness of the solution, perfect \& smooth project executions, and practicality of the project motivate participants to do their tasks satisfactorily while incomplete deliverables/ outcomes make them demotivated, resulting in losing their interest towards the project and shifting their focus to other projects. 
\par \emph{\faCommentsO ``What I am delivering at the end of the project, am I doing something useful for others and when the clients are happy with the results, I am motivated to work" -} Business Analyst (INT04)
\par \emph{\faCommentsO ``But, when you get to a point where you just had enough, and you cannot see anything completed, that we can deliver to the client, then what happen is I start working on other projects because that one is not delivering.." - } Senior Project Manager (INT12)

\par \textbf{Requirements changes} and \textbf{challenging RE-related activities} were identified as both motivating and demotivating factors for practitioners when they are involved in RE-related activities. For some of our participants (e.g., INT02, INT04, INT10, INT14, INT16, INT17, INT19, INT20, INT21), requirements changes mainly act as a motivation factor where they consider changes as challenges and doing challenging tasks keep them motivated. They emphasised that getting requirement changes enhances their involvement in the project, resulting in quality product delivery and customer satisfaction. 

In contrast, several other participants highlighted that requirements changes could be a demotivating factor for them, resulting in losing their interest towards the project. It was mentioned that constant requirements changes (specifically towards the end of the project), getting more change requests, and continuous changes in requirements in each sprint make them demotivated. It is because they feel putting lots of effort into the tasks is a waste of time due to constant changes in the requirements, creating interference in the ongoing work  (e.g., INT03, INT06, INT08, INT09, INT12, INT15, INT18). Among the practitioners who consider requirements changes as a motivating factor, the majority are business analysts or project managers, and the ones who consider it as a demotivating factor are mostly developers or team leads. 
\par \emph{\faCommentsO ``Although our client provides detailed requirements, it is good to get changes. Yes, we have to again shuffle our requirements, but it helps to get clarify the prerequisites and arrange our requirements without making any blockers to the process. Also, we get to deliver what our clients really want.." - } Project Manager (INT21)
\par \emph{\faCommentsO ``It happened towards the end of the project. We kept on getting requirements, it is like suddenly requirements are thrown at us, and it seems that they want every new idea to be included and we felt like why we put so much effort at the beginning" -}  Software Engineer (INT08)

\par \textbf{Incomplete requirements} and \textbf{poor requirements prioritization} are two demotivating factors according to the majority of our participants (18 out of 21). Incomplete requirements create uncertainty, lack of commitment, and disturbances in the development process, making the practitioners lose their interest in the work. They elaborated that it creates uncertainty in their work as they are not aware of requirements and the dependencies completely, resulting in being unable to carry forward the development work. 
\par \emph{\faCommentsO ``At the time I have been engaging with that project, they have already started it without having any proper requirements. The requirements engineering part was very poor on that project and every day was a kind of well I'm not quite sure what I'm doing, not quite sure what the team was doing, not quite sure what we are being asked to deliver"} -  Lead Human-centered Designer (INT09)
\par Similarly, poor requirements prioritization also demotivates the practitioners as it makes them feel like they are doing the wrong thing. Further, they highlighted that it might make the team frustrated and stressed, creating a bit of a chaotic situation for upcoming sprints. 
\par \emph{\faCommentsO ``Sometimes, when there's no prioritization, we get work at the end of the sprint. We get to do everything at the end of the sprint. So we get so much stressed" -}  Senior Software Engineer (INT15)

\par 8 out of 21 mentioned \textbf{technical issues} as a demotivating factor for them when doing RE-related activities. They elaborated that identifying technical mismatches with the initial design and getting lots of technical issues when discussing with the development team make them demotivated as it takes more effort to re-design and convince the customers. 
\par \emph{\faCommentsO ``We have this initial image of the application with customers, what they want to have. But when it comes to making it possible, in the technical aspects there are some mismatches and there are some things that should be included. So we have to think again, discuss with customers, it is lots of work" -}  Lead Business Analyst (INT01)

\subsubsection{Collaborative \& human Factors} \label{section 4.4.2}
We categorized factors directly related to the people involved in RE-related activities as \textbf{collaborative \& human factors}. Based on our participants' feedback, we divided this category into three sub-categories, namely, \emph{"individual"} - motivation factors related to themselves, \emph{"team"} - motivation factors related to their team members, and \emph{"customers"} - motivation factors related to their customers/ clients/ end-users. 
The majority of the collaborative \& human motivation factors are individual-related, and there are limited customer-related factors compared to both individual and team-related factors indicating that practitioners' collaborative \& human motivation factors mainly depends on themselves and their team. All the identified motivation factors in respective sub-categories and the summary of their impacts are listed in the Appendix \ref{B} (Table \ref{Table : impact of human-related motivation factors}, \ref{Table : impact of human-related motivation factors -Team}  and \ref{Table : impact of human-related motivation factors-customers}). 
The following are some original quotes for each sub-category from our interviews to illustrate these findings.

\par All of our participants mentioned \textbf{job satisfaction} as one of the main motivating factors, highlighting that the feeling of doing something useful to others and leads them to get into the details of the requirements with the desire of providing the best solution for them. This relates to the \textbf{personal satisfaction} as well because participants mentioned that apart from financial incentives, recognition, and appreciation, the self-satisfaction of doing something useful makes them motivated to perform their tasks well. For example, in INT06, a technical team lead mentioned how their job satisfaction motivates him/her as, \emph{\faCommentsO ``When I feel that I am doing something useful, and it is used by others, that keeps me doing my work. It keeps me going on and I know that I am not wasting my time. In the end, I should feel happy and satisfied with my work and I can feel it with my current work"}. 
\par \textbf{Interacting with others} is also an important motivating factor that has a positive influence on RE-related activities. According to the majority of participants (19 out of 21), it is stated that interacting with others, both internally and externally, is helpful in having in-depth discussions to elicit requirements and improves their thinking to provide advanced solutions.
\par \emph{\faCommentsO ``Interacting with others, working with people is really important to me. Especially when I work with cross-functional teams or when we have a foreign client, I am very excited to talk to them, get to know about them, you know, then I can understand surrounding, what they actually want from this project.."} - Senior Project Manager (INT12).

\par \textbf{Language \& Communication} was identified either as a motivating/ demotivating factor that impacts RE. According to INT01, INT02, INT04, INT09, INT11, INT12, and INT16, having a clear communication with others is helpful in obtaining clear descriptions of requirements for efficient work, while INT03, INT13, INT15 mentioned that the language barriers and poor communication create difficulties on understanding the problems which take additional time to complete tasks. 
\par \emph{\faCommentsO ``..we have to have good communication with the development team. It should be a two-way thing, I have a clear idea of customers' business requirements, and I should clearly communicate it to them. Also, we BAs have to be good listeners to both parties. That is really important for me.."} - Lead Business Analyst (INT01)
\par \emph{\faCommentsO ``Since I am working with clients from another country, there is a language barrier. Sometimes, specifically for clinical practitioners in that particular country, tend to use words in their own language, not the common language. Then I can't understand anything, I won't be able to get the proper picture and it frustrates me and my motivation to work on that project reduced"} -  Lead Business Analyst/Scrum Master (INT13)

\par Referring to team-related factors, having \textbf{knowledgeable, passionate, committed and responsible team members} are motivating factors for the majority of our participants, and all of these have positive influences on RE-related activities and the overall SE process. For example, 
knowledgeable team members are helpful in understanding complex requirements and dealing with the technical feasibility of the project, while having passionate, committed and responsible team members are helpful in completing the assigned tasks beyond expectations making a direct or indirect impact on the end results of the project. 
\par \emph{\faCommentsO ``One of the key things that I look for is being committed to the work rather than just doing what you are assigned to. It can be really motivating because then I can manage the requirements and timelines and everything with the stakeholders"} - Business Analyst (INT04)
\par \emph{\faCommentsO ``When everyone knows about their role and is responsible for their work, the end goal will be successful. My current team is like that"} - Business Analyst (INT16) 
\par Among the list of team-related motivation factors,\textbf{feedback from the team or customers} was identified either as a motivating or demotivating factor. 
10 out of 21 participants mentioned getting feedback either from the team or customers as a motivating factor, while some referred to it as a demotivating factor. They elaborated that it
helps them to improve user stories, prioritising the product backlog and make the progress of the project towards the right direction (INT01, INT02, INT05, INT08, INT09, INT10, INT13, INT16, INT17, INT19). However, INT04 and INT07 mentioned that they get demotivated to work when they receive negative feedback (incorrect/wrong/confusing), especially from customers and create confusion on product requirements.
\par \emph{\faCommentsO ``I do seek feedback from my manager and from the clients. But when I get some wrong feedback on my work then I get confused, I know I did my work well, but because of that I tend to think whether I understood something wrong" } - Business Analyst (INT04)

\par The participants also highlighted the \textbf{team conflicts} (15 out of 21) and \textbf{poor team management} (9 out of 21) as their demotivating factors and explained how these impact their work and the overall project. For instance, they mentioned that when they are having conflicts within the team, team members are unwilling to engage, resulting in missing requirements and taking more time to complete their tasks. Referring to the poor team management, they described it as having no clear strategy for the work, not identifying high-priority tasks and wasting resources which make difficulties in the progress of the project. 
\par \emph{\faCommentsO ``I'll choose a relatively recent example. I worked at -----six months ago and was part of their design team. We did lots of requirements gathering, actually, some of it was interface related, and some were service related. And how the management handled these were a very bad and demotivating experience for me. The primary reason for that, I think, is they do not have a clear strategy, no clear direction.."} - Lead Human-centered Designer (INT09)
\par Participants indicated several customer-related motivation factors that influence RE-related activities, and among them, almost all of the participants mentioned that having \textbf{knowledgeable customers} (20 out of 21) and \textbf{cooperative customers} (19 out of 21) make them more motivated to do RE-related activities. They explained that customers who have a clear idea of what they want and what are their priorities make them motivated to involve in RE-related activities. For example, INT05, a software engineer \& application consultant, mentioned it as, \emph{ \faCommentsO ``when the client knows exactly what he wants, that's very important from the starting point, because then we know what we have to do to make it possible and it makes our job and our life easier"}. Referring to cooperative customers, the participants elaborated that the collaboration from the clients' side, such as their involvement from the beginning to the end of the project, their involvement in decision-making, assistance in understanding the domain and providing meaningful responses to the development team inquiries motivate them to involve in RE-related activities and perform well in their tasks. For example, INT10, a business analyst, mentioned it as, \emph{\faCommentsO ``it's really important that you have good clients who are willing to cooperate with you. I mean then you will end up doing all of the requirements engineering activities well. It always helps to work with people who want to work together"}. 
Apart from these, \textbf{customer satisfaction}, \textbf{customer availability}, and \textbf{customer feedback} were the other commonly mentioned motivation factors by the interview participants, and all these factors have a positive influence on RE-related activities.
\par \textbf{Difficult to handle customers} was identified as the key customer-related demotivating factor. Participants (11 out of 21) shared that non-stop inquiries and issues from the customers' side slow down the progress of the project, and an additional pressure for them to work with narrow-minded and not engaging customers as it deviates them from focusing on the real problem and the requirements. 
\par \emph{\faCommentsO ``There were times when we had external dependencies, -- we had to wait for our clients' opinions to start the next step of the project. And our client was very strong-headed, not listening to anyone, and it was very difficult to convince him. So, every time when we had to meet him, we were exhausted at the end, and everyone in the team was like, I don't want to do anything in this project.."} - Lead Business Analyst (INT14).

\subsubsection{Socio-Technical Factors} \label{section 4.4.3}
The factors with the inter-relatedness of social and technical aspects are categorized as \textbf{socio-technical motivation factors}. 
\textbf{Domain knowledge} is identified as the most mentioned socio-technical factor 
(18 out of 21), which has both positive (motivating) and negative (demotivating) influence on RE-related activities. They also referred to it as having either the technical knowledge to make the project successful or having the specific domain knowledge to understand the business requirements. For example, some of our participants (INT05, INT06, INT07, INT08, INT10, INT14, INT15, INT18, INT19, \& INT21) explained that it is important to  have an updated knowledge of advanced techniques and technologies and be expertise in a specialized area that can be useful in the project. According to INT06, a technical team lead, practitioners with updated knowledge of advanced technologies, makes him/her motivated as it helps the team to deliver a quality outcome with less effort and time. 
\par \emph{\faCommentsO ``One of the teams that I worked in ---, we were working on projects related to financial information. The team I worked, they were always up-to-date with all these data analytic models, and everything. Highly technical team, personally I learned a lot from them and we handed over the project even before our deadline."} - Technical Team Lead (INT06)
\par Some of our participants referred to having knowledge of the business, client's needs and their background (e.g., INT01, INT02, INT03, INT07, INT11, INT12, INT13, INT20), all of which are helpful in understanding their requirements clearly. For example, INT02, a business analyst \& project manager, mentioned that working with different projects in the same industry makes him/her motivated as having prior knowledge of their business is helpful in understanding what customer needs and coming up with a solution easily. 
\par \emph{\faCommentsO ``I did 4-5 projects in the same industry, so I have a better understanding of their requirements. That's my speciality area"} - Business Analyst \& Project Manager (INT02)
\par Further, two of our participants (INT07 \& INT10) highlighted that having less knowledge of their project domains demotivated them as it took more time and effort to understand the requirements, and they faced difficulties in coming up with better solutions. INT10, a business analyst who is currently working on projects related to utility services, mentioned that the current projects that she/he is part of are making her/his demotivated as it is hard to understand the domain and as a result, she/he is facing difficulties in understanding the requirements and conveying them to the development team. 
\par Having a \textbf{challenging problem to solve} is another motivating factor, according to the majority of our participants (17 out of 21). They referred to it as either having a challenging task to complete in each sprint or having a challenging business problem that needs to be given a technical solution. Among these 17 participants mentioned that having challenging problems to solve motivates them when doing RE-related activities as it gives them the opportunity to learn something new, do experiments with unfamiliar technologies, break down a massively challenging problem into doable tasks and come up with solutions to complex requirements resulting in obtaining clarity of requirements and self-satisfaction. 
\par \emph{\faCommentsO ``I like to get into the challenging tasks where people got stuck and solve it and implement a solution for it. It's motivating for me"} - Technical Team Lead (INT06)

\par However, some of our participants mentioned that having a challenging problem to solve demotivates them when involved in RE as it distracts their focus on the project. They further elaborated that having a challenging problem requires lots of time and effort, and in the end, they get bored with providing the best solution for it.  For example, INT19, a business analyst explained it as, \emph{\faCommentsO ``I can be easily distracted when I'm having a challenging, complex problem. In my last project, I had to put lots of effort into understanding its initial days, and I felt like I'm wasting my time. There was no progress. Because of that, I was down and thanked our project manager and the team, we completed it"}. \\ 

\textbf{Project diversity} is identified as another socio-technical motivating factor. 
Over half, 13 out of 21 participants, mentioned that they get motivated when they get to work on diverse projects and referred to it as either working on projects from diverse domains or using diverse techniques for RE-related activities or getting a variety of tasks to complete in each sprint. 
They further elaborated that working on different domains and using different techniques to gather requirements are helpful in keeping their focus on the project while getting the opportunity to try and learn new things. \\
 \emph{\faCommentsO ``In different projects, I use different techniques to gather requirements. 
... Sometimes you get the requirements very easily and sometimes we have to put in lots of effort. So, I have to use various techniques accordingly, such as interviews, brainstorming sessions, and workshops. Through these years, I learned these things, and now I get very excited when I get to use them and share my knowledge with juniors"} - Lead Business Analyst (INT01) 
\par Apart from these, \textbf{task workload}, \textbf{project timeline}, \textbf{cross-functional work} and \textbf{following standard process} were also identified as socio-technical motivation factors as participants referred to these factors from both technical and social perspectives. For example, INT10, a business analyst, mentioned that having a proper project timeline motivates them as it is a deciding factor of how much effort she/he should put into each allocated task, whereas according to INT09, a lead human-centred designer, it is an additional pressure to work for a timeline, especially when there are strict deadlines. Similarly, for task workload, some participants (INT04, INT05, INT10, INT15, INT17, INT18, INT19) consider it as a motivating factor, whereas for some (INT14, INT16, INT20), it is a demotivating factor. For instance, according to INT10, a business analyst, \emph{ \faCommentsO ``in requirements engineering, when there are more workshops, you have to do lots of work, there are more discussions, but it's motivating to work because then you get a better understanding and do your work well"}. Cross-functional work is mentioned as a motivation factor by 9 out of 21 participants, who explained that though it is challenging, it gives them a clear idea of what they expect from each and every function, making the implementation easier. 
\par \emph{\faCommentsO ``There's been a lot of consultation with various teams as the project is going across the AAA. We have to spend a lot of time with each and every team to come up with a central process. It has given us lots of clarity of how the process should work for everyone"} - Lead Business Analyst \& Product Owner (INT17).

\subsubsection{Social Factors} \label{section 4.4.4}
Based on our analysis, we identified a few factors related to the social context as motivation factors for practitioners when involved in RE-related activities. 14 out of 21 participants mentioned at least one of these factors, and we categorized them as \textbf{social motivation factors}, namely \textbf{benefit for the society}, \textbf{working culture} and \textbf{gender biases}. Among these, the benefit for the society was identified as a motivating factor with positive influence, whereas working culture and gender biases were identified as factors with both positive and negative influence on RE-related activities (both motivating \& demotivating).
Referring to the \textbf{benefit for the society}, 12 out of 21 participants mentioned that the project's impact on society, how it makes the end user's life easier, and the thought of someone's life gets impacted by the project make them motivated to do RE-related activities successfully. They elaborated that when the project benefits society or a larger population, it motivates them to do their work better as they want to come up with the best solution possible. 
\par \emph{\faCommentsO ``For me, one of the biggest motivations is to think about the impact it's doing on society. I think about how it's changing people's lives, how it's giving benefits to everyone from the product. So, I try my best to do my part in the best possible way"} - Software Engineer (INT07)
\par 9 out of 21 participants mentioned that \textbf{working culture} as a motivation factor for them to do RE-related activities, and based on the influence, it can be either a motivating/demotivating factor.  According to our participants INT07, INT08, INT10 and INT11, INT16, INT19, when they are having a fun working environment, a good team culture with a friendly environment, and an open, diverse work environment facilitated with everything, they get motivated to do their best in every allocated task resulting in getting the best outcome at the end. However, for some participants (INT05, INT14, INT18), the working culture can be a demotivating factor when they have to work in a less supportive team environment or get the work done by interacting with lots of culturally different people. 
\par \emph{\faCommentsO ``There were 20-25 members in that team, it was a big project, like two internal teams, and two external teams from Sweden. The working culture is very different and I had to interact with all of them and sometimes it was really tough.."} - Lead Business Analyst (INT14)
\par Although the majority of the participants did not mention \textbf{gender bias} as a motivation factor, 6 out of 21 mentioned that gender bias at work is real and it is a motivating or a demotivating factor for them when involved in RE-related activities. For the majority, it is a demotivating factor, but some of our participants take it as a motivation to do their best. According to INT02, a business analyst \& project manager, it influences positively as it makes them do their work best. \emph{\faCommentsO ``Yes, I have experienced it. There was this huge discrimination in one of my past projects. But I didn't feel bad about it. I got motivated, and in the end, I did better than others"} - Business Analyst \& Project Manager (INT02). As mentioned, for the majority, it is a demotivating factor, resulting in taking more time to complete the tasks. \emph{\faCommentsO ``All the places I worked, there was a heavy tendency for male developers. I think they were somewhat reluctant to work with a female BA. So I also felt somewhat hesitant to give my suggestions or talk about changes at first..."} - Business Analyst (INT10).
\par Referring to all the identified motivation factors, the participants explained how these factors influence RE-related activities, supporting the phenomenon of our theory. Further, from all the identified motivation factors, the majority are collaborative \& human motivation factors. This indicates that people are an integral part of software development, and from the practitioners' perspective, \emph{motivation} is considered to be one of the largely impacted human aspects on RE. 

\subsection{Strategies - Action/Interaction Strategies to Overcome, Handle or React to the Influence of Motivation on RE-related activities} \label{section 4.5}

Strategies in the coding paradigm define as \emph{the action or interaction strategies that are directed towards the phenomenon irrespective of whether the research is focused on individuals, groups or collectives that can take place to overcome, handle or react to the phenomenon} \cite{strauss1990basics}. In our study, we identified 12 strategies that can be used by practitioners to reinforce the positive influences of motivation on RE-related activities while overcoming the negative influences of demotivating factors (figure \ref{Theory}). 

\par \textbf{S1. Exposure to experiments:} The most common strategy mentioned by almost all the participants (20 out of 21) is providing diverse, challenging tasks to the practitioners to keep their motivation throughout the project. They shared that with the diverse and challenging tasks, they get the exposure to do experiments and learn new things resulting in completing each allocated task best possible way. For example, INT14, a lead business analyst, highlighted that when doing requirements elicitation, it is important to give opportunities to the team to try out new techniques rather than just doing discussions and workshops with the clients as it makes the team members motivated to involve in the meetings with the clients. \\
 \emph{\faCommentsO ``So, this project was for --- hospital, we had to interact with healthcare workers, from top management to the ones who handle patients. It was very challenging, --with all the medical terminologies and processes. So, I thought of trying multiple options and let other BAs in the team to come up with various ways that we can have in these client meetings to get the requirements. The BAs got really excited, I'd say motivated after that because it was fun for them, combining UIs with stories, simplifying each process, at the end everyone even the client side was very excited.."} - Lead Business Analyst (INT14)

\par Further, they elaborated that providing proper exposure to do experiments and learn is helpful in their career development, especially when they are at the early stage of their career. As a result, when they are given the opportunity and space to learn and utilise that learning in practice, they get highly motivated to involve in RE-related activities. 

\par \textbf{S2. Rebuild collaboration:} 20 participants mentioned the importance of having good collaboration among their team members and also between the development team and the customers. They highlighted that when the team or the customers are not engaging due to various reasons, such as conflicts, everyone involved in the project gets demotivated. One of the first steps they take to overcome that situation is trying to rebuild that collaboration by changing responsibilities in the team, initiating supportive online channels (slack/email threads) with the team by identifying everyone's strengths \& weaknesses, and having individual or group discussions with the team and the clients.\\ 
 \emph{\faCommentsO ``The way in which I turned that around was a kind of a bit different. It's not that everyone in the team was not happy to engage, only some of them. First I got an idea of who are they, identified who are they comfortable working with, and created supportive channels where they can get together and work first. Initially, it was like for a small group in the team like two members working together, slowly I made it to a bigger group.."} - Lead Human-centered Designer (INT09)
\par Participants INT01, INT09 \& INT10 also highlighted that mutual trust and time are two vital factors that are needed to rebuild the collaboration either between team members or with clients. They further elaborated that when there is mutual trust and availability to spend some time to understand others, it is easier to rebuild the collaboration. 

\par \textbf{S3. Open \& individual discussions:} Formal, informal, open, one-on-one, all sorts of discussions are important in keeping the team motivated and overcoming demotivated situations within the team or with clients. 19 out of 21 participants explained that they use one or more forms of these discussions as a strategy to handle motivation. Using sprint retrospectives to conduct transparent discussions, quick informal discussions with multiple members to develop mutual understandings on doing sprint tasks collectively, and arranging one-on-one meetings among the manager \& members or with the clients is helpful in making everyone in the team well-informed. These discussions are helpful in sorting out the differences and getting everyone's point of view on how to complete the tasks considering the dependencies.  
\par \emph{\faCommentsO ``We try to sort out issues by being open about it. We discuss all these in retros and if something becomes urgent, we set up a quick meeting with the responsible ones to sort it out. It is very important because then everything is clear and we can focus on our work.."} - Lead Business Analyst \& Scrum Master (INT13 )

\par \textbf{S4. Focus on task completion:} Irrespective of the participant's role in the project, most (17 out of 21) mentioned that whenever there is a demotivating situation,  to overcome it they tend to only focus on completing their allocated work within their own pace of work.
\par \emph{\faCommentsO ``At the end, none of us really cared, because we couldn't handle that situation. So we were just sitting around doing whatever was allocated to us, nothing else"} - Senior Business Analyst (INT20)
\par Further, they elaborated that turning a demotivated team into a motivated team is far more challenging than keeping a motivated team as it is. Hence, it is always good to let the team slow down the progress and try to complete the tasks as much as they can. For example, INT12, a senior project manager explained that it is okay to let the BAs (business analysts) gather requirements the way they want, not expecting to put extra effort into getting detailed requirements. \\
\emph{\faCommentsO ``I didn't worry about getting all the requirements we need. I knew the situation and that sometimes you should allow time to solve problems. So I thought it's better to work on something rather than nothing and let the BA do her work and the dev team do their work..} - Senior Project Manager (INT12)

\par \textbf{S5. Flexible task delegation:} One of the most commonly mentioned strategies (15 out of 21 participants) is flexible task delegation. They elaborated that although the highly prioritized tasks are assigned to the expertise in that particular area, it is important to make the task delegation flexible for the team members so that they may not feel bored or demotivated by getting similar kinds of tasks. 
\par \emph{\faCommentsO ``I'd say, the motivation for us is to get that chance to pick up the particular task in each sprint, so we can really work on whatever we want to work. It's our manager who looks into this making sure that we have that autonomy while getting an equal contribution from all of us"} - Software Engineer (INT08)
\par However, some also highlighted that in projects where they require specific technical skills, using the strategy of flexible task delegation to keep the team motivated can be a limitation. 

\par \textbf{S6. Being empathetic:} is another strategy that can be used to overcome a demotivated situation when involved in RE-related activities. 15 out of 21 explained that being empathetic is important when involved in RE-related activities as it is the phase where they have to engage with clients, end users, developers, and testers all together and deal with their emotions, anxiety, and personalities while doing RE-related activities. Hence, it is essential to understand that one demotivated person can create a whole demotivated situation and empathize with them can overcome these situations because then the practitioners tend to pay more attention to observing and understanding what they really want.  
\par \emph{\faCommentsO ``We have to deal with lots of emotions of people. So we have to understand that every day is not the same and we can't expect everyone in the team to work the way we expect. I always try my best to cheer them up when I see that they are feeling low. I pay more attention to understand their situation, if I can provide some advice because I don't want them to be in that zone for a long time.."} - Software Engineer \& Application Consultant (INT05)

\par \textbf{S7. Changing the process or people involved:} 12 out of 21 participants claimed that changing the process or the people is a strategy they use to avoid demotivated situations when doing RE-related activities. Referring to the process, it can be done by using different approaches to complete tasks, following certain pre-defined practices according to situations, changing the type of meetings \& discussions by assessing the ongoing process for elicitation, re-prioritizing the requirements or coming up with a rolling period for re-organizing the allocated work whereas referring to the people, it can be done by adding more resource persons to handle RE-related activities, changing developers based on the expertise or providing the opportunity to change for less contributing individuals. \\
 \emph{\faCommentsO ``What I'll do is I'll properly observe how we collected the requirements, what is missing and we have certain practices that we follow. So we know what type of meeting is needed, with whom we have to arrange it to get the work done.."} - Business Analyst \& Project Manager (INT02)
\par \emph{\faCommentsO ``When one can't find answers, I'll tell someone else to carry forward that task. But before that, I make sure they are ok with it, balancing their work. We have to understand that everyone in the team is not an expert in dealing with requirements"} - Lead Business Analyst \& Scrum Master (INT13)
\par \textbf{S8: Seek management support:}
Half, 12 out of 21 participants, also stated that seeking management support when needed is one of their strategies to deal with demotivating situations. They specifically highlighted that they use this strategy in situations where support from the management is necessary to take the team out of that demotivating situation. They explained that project managers/ leads should have a proper idea of when to raise the concern to the management, what to agree and what not to agree on in their solutions without making the situations more complicated. Further, they provided examples of those situations such as unrealistic client requests, complicated team conflicts, and exceeding the project timeline or the budget that can make the whole team demotivated to work.
\par \emph{\faCommentsO ``Usually, when we have daily stand-ups, we discuss our problems every morning. But sometimes we have to face things that we didn't expect, like some change requests, no one in the team is willing to do because we know it'll take lots of time and effort. So, it's kind of bothering us because, from the client's side, it's difficult to manage. Everyone is waiting for the managers to solve it.."} - Senior Software Engineer (INT15)

\par \textbf{S9. Providing regular updates:}
11 out of 21 participants mentioned that it is important to provide regular updates to make everyone in the team informed and avoid miscommunications and misunderstandings that can create demotivating situations. They further elaborated that when there are meetings with the clients and key team members who are involved in RE, it is necessary to keep the whole team updated with proper communication to sort out issues in requirements so that the implementation can be done without any disturbances. 
\par \emph{\faCommentsO ``I make sure that everything is sorted and everyone is updated, like all the team members, so that they can do their work without any issue.."} - Business Analyst \& Project Manager (INT02)
\par Also, they explained that it is helpful to focus on prioritized tasks without any conflicts or complicated situations that can make the team demotivated, and these regular updates are very important, specifically when the team size is big. \\
\emph{\faCommentsO ``When the team size is so big, we can't take everyone for the requirements engineering phase. We try to limit some team members and take key team members to client meetings. So we specially make sure that everyone is updated at the end of the day so that there won't be any conflicts.."} - Project Manager (INT21)

\par \textbf{S10. Equal opportunity \& freedom to speak:} A small number, 6 out of 21 participants, mentioned a few instances where they considered providing equal opportunity \& freedom to speak as a strategy to keep the team motivated. They further explained that this strategy is useful when analysing and prioritizing the requirements with the team in sprint planning and story walk-through sessions so that the team members can express their opinions without any hesitation and helpful in getting a clear perspective of the expected outcome.\\ 
 \emph{\faCommentsO ``Sometimes it will be kind of a primary level question, but I make sure to give value to each question. Whether it's with the clients or dev team, I listen to them, and let them speak, so that I get to know what they think. In a way it is important to prioritize our work in each sprint and in the team, they will feel involved when taking decisions.."} - Lead Business Analyst (INT01)

\par \textbf{S11. Making deadline-driven projects:} A few participants (INT02, INT14, INT15 \& INT18) mentioned that making the projects/tasks deadline driven is helpful in keeping the motivation within the team. They described that a properly planned deadline for each task in the sprint is useful in keeping the team focused and on track. \\
 \emph{\faCommentsO ``what I do is, try to keep the focus on track. Actually,  I learned this from my previous project manager. I have a deadline for each task, small ones, big ones, not just random deadlines, I think about the tasks and then decide. With that I can motivate the team and feels like we all work for a common goal"} - Business Analyst \& Project Manager (INT02)\\
 However, INT18, a software engineer, elaborated that although having deadlines motivates them to work, there should be the ability to push deadlines sometimes to mitigate the stress. \\
\emph{\faCommentsO ``They should be able to push that deadline from like a couple of days. That actually worked for us and helped us to mitigate our stress and also be motivated"} - Software Engineer (INT18)

\par \textbf{S12. Moving out from demotivated situations:}
A few of our participants (4 out of 21) stated that moving out from demotivated situations is one of their strategies to overcome demotivating situations.  They further elaborated that when there are unavoidable conflicts within the team that cannot be solved, or when there is no meaningful work to do, they do not prefer to be there as it demotivates them. In those scenarios, they prefer moving out from the allocated tasks, projects or even from the company to get over the demotivating situations. \\
 \emph{\faCommentsO ``I really had to move out from that project, because there were many conflicts in the team, the management was not giving any solution and I decided it's better to back off and focus on other projects, to give my 100\%.."} - Lead Business Analyst (INT14).

\begin{table*}[htbp]
\caption{Consequences of applying strategies to manage the influence of motivation on RE (*Conseq-Consequences; *R-S - Related Strategies) }
\label{Table: consequences}
\Large
\resizebox{\textwidth}{!}{%
\begin{tabular}{@{}clll@{}}
\toprule
\textbf{*Conseq.}                                 
& \textbf{Description}                                            
& \textbf{Original Quotes} 
& \textbf{\begin{tabular}[c]{@{}l@{}} *R-S\end{tabular}} 
 \\ \midrule

\begin{tabular}[c]{@{}l@{}}C1. Product\\ delivery    \end{tabular} 
& \begin{tabular}[c]{@{}l@{}} Obtaining the final product\\ aligned with customer expec\\-tations, correct specifications  \\resulted in customer satisfaction, \\generate good revenue \& increase \\company reputation\\ (19/21 participants) \end{tabular}            
& \emph{\begin{tabular}[c]{@{}l@{}} \faCommentsO ``we could deliver the software with correct specifications,\\ and I think it's because we worked together to make\\ it happen"- INT06 (Technical Team Lead) \\  \faCommentsO``At the end what matters is the customers are happy and \\satisfied with our work. It's very important to our \\reputation" - INT04 (Business Analyst) \\ \faCommentsO  ``We need to focus on getting all the requirements and I \\make sure that it should happen, I don't mind about the \\process, what it matters is the perfect outcome" \\- INT14 (Lead Business Analyst) \end{tabular}  }
& \begin{tabular}[c]{@{}l@{}} S2, S3,\\ S7, S8\end{tabular} 
\\  \cmidrule(l){1-4} 

\begin{tabular}[c]{@{}l@{}}C2. Effective\\ process  \end{tabular} 
& \begin{tabular}[c]{@{}l@{}} Have fun and effective process to \\follow without flaws. For example,\\ there will be fewer disruptions in\\ the process, due to demotivating\\ situations  which is helpful in \\completing tasks well \\ (17/21 participants)\end{tabular}            
& \emph{\begin{tabular}[c]{@{}l@{}} \faCommentsO ``Either you have to change the process or the people \\involved in that process to make it work. Either way, finally\\ we get what we need without any trouble.." - \\INT13 (Lead Business Analyst \& Scrum Master) \\\\ \faCommentsO ``I try to allocate tasks according to everyone's preference,\\ most of the time I let them pick whatever they want, but \\make sure everyone has to put the same effort. So that \\everyone is happy and at the end, we'll have productive sprint.." \\- INT06 (Technical Team Lead) \end{tabular}}
& \begin{tabular}[c]{@{}l@{}} S5, S7, \\S11 \end{tabular} 
\\  \cmidrule(l){1-4} 

\begin{tabular}[c]{@{}l@{}}C3. Fewer\\ conflicts  \end{tabular} 
& \begin{tabular}[c]{@{}l@{}}  Less internal \& external conflicts.\\This can be due to proper collabo\\-ration among the team, empathizing\\ towards the clients, getting opportu\\-nity to talk to solve problems and \\support from the management when\\ needed (17/21 participants)\end{tabular}            
& \emph{\begin{tabular}[c]{@{}l@{}}  \faCommentsO ``you know, whatever reasons, I'd have a conversation with \\them individually to understand their point of view, to figure \\why they do not like our suggestions without having any issue \\with them.." -INT04 (Business Analyst) \\\\ \faCommentsO ``I want people to express explicitly because I'm open in\\ discussions, like talk openly. I believe that's the best way to \\resolve those conflicts" - INT14 (Lead Business Analyst) \end{tabular} }
& \begin{tabular}[c]{@{}l@{}} S2, S6, \\S8, S10, \\S12\end{tabular}  
\\  \cmidrule(l){1-4} 

\begin{tabular}[c]{@{}l@{}}C4. Satisfaction  \end{tabular} 
& \begin{tabular}[c]{@{}l@{}} Improve personal satisfaction of the\\ practitioners, satisfaction as a team \\after completing the project and\\ obtaining customer satisfaction at\\ the end (16/21 participants) \end{tabular}            
& \emph{\begin{tabular}[c]{@{}l@{}} \faCommentsO ``In the end, I should feel happy and satisfied with my work\\ and I can feel it with my current work" \\-INT06 (Technical Team Lead)\\\\ \faCommentsO ``It was a very difficult situation to manage the customers \\because there were some delays from our end. So we made sure\\ to work on a strict deadline for couple of sprints to deliver what \\they want to, and at the end, they were satisfied with our work.."\\ -INT15 (Senior Software Engineer) \end{tabular}  }
& \begin{tabular}[c]{@{}l@{}} S1, S4, \\S5, S10, \\ S11\\ \end{tabular}  
\\   \cmidrule(l){1-4}  
 
 \begin{tabular}[c]{@{}l@{}}C5. Individual \\ performance \end{tabular} 
& \begin{tabular}[c]{@{}l@{}}  Improve individual performance \\of the practitioners by getting \\the opportunity to learn, work on\\ desired tasks \& with others, and\\  getting away from situations that can\\ slow down their progress\\ (14/21 participants) \end{tabular}            
& \emph{\begin{tabular}[c]{@{}l@{}} \faCommentsO ``I would let them think and take the decisions by themselves\\ first, otherwise they won't learn how to overcome that situation.\\ Sometimes, it's a kind of teaching, because for any project there \\will be difficult situations and we must be prepared. When I \\see that they are doing it well, I also feel motivated.." \\- INT09 (Lead Human-centered Designer) \\ \\ \faCommentsO ``Finally, we resolved all the issues with the clients and we \\all were on the same page. Then everyone put their effort into\\ making the project successful and personally I think that was one \\of my projects where I performed really well.."\\ - INT20 (Senior Business Analyst) \end{tabular}  }
& \begin{tabular}[c]{@{}l@{}} S1, S2,\\ S5, S11,\\ S12 \end{tabular}  
\\  \cmidrule(l){1-4} 

 \begin{tabular}[c]{@{}l@{}}C6. Product \\ Quality \end{tabular} 
& \begin{tabular}[c]{@{}l@{}}  Improve the quality of the product \\with improved features beyond \\customer expectations, satisfying\\ all of their requirements and\\ in-line with market trends \\(15/21 participants) \end{tabular}            
& \emph{\begin{tabular}[c]{@{}l@{}} \faCommentsO ``At the end, we provided a quality outcome. It was the best\\ one to use as the real time ERP system. We completely automated\\ the system and it was a success because everyone was motivated \\and did lots of hard work.."\\ - INT03 (IT Development \& Re-engineering Leader)\\\\ \faCommentsO ``Because we got all the little details required to complete \\the project, we could provide the best solution to our customers..." \\- INT21 (Project Manager)  \end{tabular}  }
& \begin{tabular}[c]{@{}l@{}}S1, S3, \\ S7, S9\end{tabular}  
\\   \cmidrule(l){1-4} 

 \begin{tabular}[c]{@{}l@{}}C7. Customer \\ engagement\end{tabular} 
& \begin{tabular}[c]{@{}l@{}} Increase the customer involvement\\ with proper collaboration, individual\\ \& open discussions, updates and\\ even with management support result\\-ing in obtaining clear complete \\requirements \& reduce last \\minute changes (14/21 participants)\end{tabular}            
& \emph{\begin{tabular}[c]{@{}l@{}}\faCommentsO ``we had a very good relationship with our clients. It's because\\ from day one we were open about everything, and when there\\ is an issue, disagreement,  we directly discussed it with them.\\ So I think that helped us a lot dealing with changes" \\- INT10 (Business Analyst) \\\\ \faCommentsO ``We make sure to keep them [clients] updated to get their \\involvement in each stage. This way we won't get any last-minute\\ surprises from their end" \\- INT01 (Lead Business Analyst) \end{tabular}  }
& \begin{tabular}[c]{@{}l@{}} S2, S3, \\ S8, S9 \end{tabular}  
\\   \cmidrule(l){1-4} 

 \begin{tabular}[c]{@{}l@{}}C8. Timeliness\end{tabular} 
& \begin{tabular}[c]{@{}l@{}} Completing and delivering projects/\\ products on-time resulting in\\ customer satisfaction, good\\ reputation of the company, and no \\exceeding budget limits due to\\ surpass timelines (12/21 participants) \end{tabular}            

& \emph{\begin{tabular}[c]{@{}l@{}}\faCommentsO ``No matter what happens, I wanted to complete the \\project because it was our initial days as a company. So, I \\made some tough decisions, put some strict deadlines to\\ make it happen and finally, we delivered the project on time" \\- INT07 (Software Engineer) \end{tabular}  }
& \begin{tabular}[c]{@{}l@{}} S4, S5, \\ S7, S11 \end{tabular}  
\\   \cmidrule(l){1-4} 

 \begin{tabular}[c]{@{}l@{}}C9. Quality \\ RE\end{tabular} 
& \begin{tabular}[c]{@{}l@{}} Getting complete, detailed require\\-ments of the project at the early \\stage and manage requirements \\throughout the project easily  \\(15/21 participants) \end{tabular}            
& \emph{\begin{tabular}[c]{@{}l@{}} \faCommentsO ``We'll have all the required people in one meeting, from\\ tech leads to product owners and everyone so that we can \\discuss all together in one meeting. It really helps to get through\\ challenging requirements"\\ - INT02 (Business Analyst \& Project Manager) \\\\ \faCommentsO ``I always assess the process to make sure everyone is on\\ the same track, specially in first couple of sprints. When I \\feel that something is not going the right way, I try to use \\alternative ways to get them on track, because my ultimate\\ goal is to get do quality work"\\ - INT16 (Business Analyst) \end{tabular}  }
& \begin{tabular}[c]{@{}l@{}} S1, S2,\\ S3, S6,\\ S7, S9\end{tabular}  
\\  \cmidrule(l){1-4} 

 \begin{tabular}[c]{@{}l@{}}C10. Individual \\ improvement\end{tabular} 
& \begin{tabular}[c]{@{}l@{}}  Individual improvement of \\software practitioners through learn\\-ing and getting exposure to new \\techniques and technologies, and \\capability of handling situations \\with the focus of completing the\\ allocated tasks (11/21 participants) \end{tabular}            
& \emph{\begin{tabular}[c]{@{}l@{}}\faCommentsO  ``I am particular about getting everyone's opinion,  I tell\\ them that leads can also do mistakes, so if you feel it's not\\ correct, tell me about it. So, they get that confidence to speak,\\ and I believe we should give that experience to them" \\- INT09 (Lead Human-centered Designer) \end{tabular}}  
& \begin{tabular}[c]{@{}l@{}}S1, S3, \\S4, S10 \end{tabular}  
 \\  \cmidrule(l){1-4} 
\end{tabular}%
} 
\end{table*}

\subsection{Consequences - of Applying Strategies to Manage the Influence of Motivation on RE-related activities} \label{section 4.6}

Finally performing these action/interaction strategies lead to consequences. 
\emph{The consequences can be real or hypothetical in the present or in the future, and in one phenomenon, these can be consequences of an action/interaction strategy, whereas, at a later point in time, they can be part of causal conditions for another phenomenon \cite{strauss1990basics}}. In our study, we define these consequences as the impact of adopting these strategies to manage the influence of motivation on RE-related activities. As shown in figure \ref{Theory}, we identified 10 main positive consequences and table \ref{Table: consequences} explains the consequences of applying identified strategies supported with original quotes from the interviews. 
\par These findings suggest  that motivation plays an important role in doing RE-related activities successfully and delivering a quality outcome at the end. However, these identified factors that motivate or demotivate software practitioners, strategies and consequences can vary on the contextual conditions, and as a result, the influence on RE-related activities can also vary, which requires further investigations.

\section{Discussion} \label{section 5}

\subsection{Key findings} \label{5.1}
The key finding of our study is the theory (Figure \ref{Theory}) on how motivation influences RE-related activities. We discovered that motivation plays a crucial role in successfully conducting RE-related activities, with its impact varying based on contextual conditions such as project domain, practitioners' experience, job roles, involvement in RE activities, and the software development methodologies used (sections \ref{section 4.2} and \ref{section 4.3}). Thus, our findings are particularly relevant for understanding the influence of motivation on RE activities within an agile software development context.

We identified that software practitioners are motivated by various factors when involved in RE-related activities, which can be either motivating, demotivating, or both (section \ref{section 4.4}). These factors are categorized into four main groups: \emph{technical}, \emph{collaborative \& human}, \emph{socio-technical}, and \emph{social}. The collaborative \& human category is further divided into \emph{individual}, \emph{team}, and \emph{customers} sub-categories (section \ref{section 4.4.2}). We found that individual and team-related factors are more prevalent than customer-related factors, indicating that software practitioners and their teams play the most crucial role in either facilitating or hindering motivation's impact on RE. These factors influence not only RE-related activities but also the overall development process and the final project outcome.


\par We also identified 12 strategies that are used by practitioners to manage the positive influences on motivation during RE-related activities wand overcome demotivating factors (section \ref{section 4.5}), along with the consequences of applying these strategies to manage the influence of motivation on RE-related activities (section \ref{section 4.6}).  
However, according to our participants, the use of these strategies may vary based on their organization, the projects, their role on the project and the people they work with. For example, some strategies (S1, S7, S10, S11) can only be applied by the team lead or project managers, whereas some of the strategies (S8 -seeking management support \& S12 - moving out from demotivated situations) have been used in rare circumstances that have happened only once or twice as of their experiences.  Interestingly, almost all the participants stated that they are highly motivated to work on their current projects, and the strategies they mentioned are helpful in keeping the motivation when they involve in RE-related activities. 

The consequences identified in this study are the positive outcomes of applying the strategies to manage the influence of motivation on RE-related activities, and one consequence can occur with the use of one or more strategies. For example, applying various strategies such as S3 - open \& individual discussions, S6 -being empathetic, or S9 - providing regular updates can keep the practitioners motivated, resulting in getting complete, detailed requirements (C9 - Quality RE - Table \ref{Table: consequences}). Further, our findings suggest that the team leads (lead business analysts, technical leads) and project managers play a significant role in keeping everyone in the team motivated while overcoming demotivated situations via using these strategies that impact RE-related activities, final outcome of the project and the people who are involved in it.

\subsection{Key Insights \& Reflections} \label{section 5.2}


\par \textbf{Impact of the context:} We identified that context plays an important role in applying our theory  of understanding the influence of motivation on RE-related activities. Therefore, the motivating or demotivating factors we identified from our interview analysis may depend on participants' experience and involvement in RE-related activities, their role in the project, and the domain of the project they referred to. \emph{"Job satisfaction"} was identified as collaborative \& human motivation factor mentioned by all the participants. However, the reasons they provided for why job satisfaction is a motivating factor for them vary based on their project domain. For example, participants involved in the health or education domain highlighted that the feeling of doing something for the end users (e.g.patients or students) makes them motivated to do their tasks, whereas others referred it as the feeling of completing their tasks successfully, resulting in quality output. Similarly, participants who highlighted   \emph{"career development"} as their motivation factor mostly have less than 10 years of experience in the field. This highlights the need for further investigation of the impact of these contextual conditions when focusing on motivation or other human aspects in future studies. 

\par \textbf{Relationship between personality and motivation:} We identified that there might be a relationship between individuals' personalities and the factors that motivate practitioners when they are involved in their RE-related tasks. From the motivating/ demotivating factors we identified under the individual sub-category of collaborative \& human motivation factors, factors such as hard-working nature, supportive nature, interaction with others, curious nature, empathising with others, willingness to explore directly relate to characteristics that are being used to identify personality traits in the standard five-factor model \cite{RN2997}. For example, hard-working nature relates to the "conscientiousness" trait, curious nature relates to the "extraversion" trait, whereas the willingness to explore relates to the "openness to experience" trait, indicating that individual-related motivation factors can depend on the practitioners' personality as well. This insight opens up future research into focusing on incorporating personality and motivation aspects when performing RE-related activities. 

\par \textbf{Impact of external forces:}
We identified that external forces influence practitioners' motivation when involved in RE-related activities. Among them, customer-related factors play an important role as knowledgeable customers, cooperative customers, customer satisfaction, customer availability and their feedback are considered as motivating/ demotivating factors by almost all participants (20 out of 21). Further, social motivation factors such as working culture was also mentioned by the participants as a motivating factor when involved in RE-related activities that described as fun, friendly, diverse working environment positively influenced their motivation, whereas less supportive working environment demotivated them when doing RE-related activities. Hence, future studies can explore how the role of customers and working culture impact RE-related activities.

\par \textbf{Perspectives of practitioners from different vs same team:} As we conducted in-depth interviews in two iterations, the participants in our second iteration belonged to the same software development team, working on the same project in the education software domain in Australia. Participants in the first iteration are from various software development teams from various countries. Thus, we identified some similarities as well as differences in certain motivating/ demotivating factors and strategies they used when involved in RE-related activities. For example, in the first iteration, cross-functional work was mentioned as a motivation factor by only INT03 and INT09, whereas almost everyone in the second iteration mentioned it as a motivation factor for them as they were involved in gathering requirements from cross-functional teams in their ongoing project. Similarly, "gender bias" was mentioned as a demotivating factor by six participants and among them, five of them were from iteration one, whereas only one participant (INT19) mentioned that he/she experienced gender bias and it is a demotivating factor for him/her. However, he/she also emphasised that he/she experienced it in his/her previous working place and there is no gender bias in the current team he/she works with. Also, when referring to the strategies they used to keep their motivation, exposure to experiments (S1), open \& individual discussions (S3), flexible task delegation (S5) and providing regular updates (S9) were highlighted by the majority of the participants in the second iteration. This indicates that participants' perspectives and experience on the influence of motivation on RE-related activities can vary based on the team they work with. It suggests the need for more follow-up clarifications by observing different software development teams. 

\par \textbf{Emergence of the social category:}
In terms of categorization, it is difficult to group some factors into one particular group. For example, when categorizing factors such as working culture and gender biases, it can be related to collaborative \& human categories as well. However, when analysing the data, we identified that participants referred to it from a social perspective. For example, when referring to working culture, the participants' opinions were more towards the friendly working environment or cultural diversity of the working environment, whereas when referring to gender bias, they talked about it as social discrimination. Therefore, we categorized these as social motivation factors. However, further studies can be designed to capture these factors better, considering the perspective of the wider population of software practitioners. 

\begin{table*}[]
\caption{Comparison of the study with the existing related work}
\label{Table: comparison with related work}
\Large
\resizebox{\textwidth}{!}{%
\begin{tabular}{@{}cllll@{}}
\toprule
\textbf{Study}          
& \textbf{\begin{tabular}[c]{@{}l@{}}Context \\of study\end{tabular}}                                     
& \textbf{\begin{tabular}[c]{@{}l@{}} Method/\\Approach\end{tabular}} 
& \textbf{\begin{tabular}[c]{@{}l@{}} Motivating factors \\identified\end{tabular}} & \textbf{\begin{tabular}[c]{@{}l@{}}De-motivating factors \\identified \end{tabular}} \\ \midrule

\begin{tabular}[c]{@{}l@{}}Beecham et al.\\ \cite{BEECHAM2008860} \end{tabular}
& \begin{tabular}[c]{@{}l@{}}  Motivation in \\SE; focusing \\on what moti\\-vate/demotivate \\developers\end{tabular} 
& \begin{tabular}[c]{@{}l@{}} An SLR \\- analysing \\92 papers \end{tabular} 
& \begin{tabular}[c]{@{}l@{}} 21 motivators - Rewards \& \\Incentives, Development needs \\addressed, Career path, Variety \\of work, Good management, Work-\\life balance, Feedback, Recognition,\\ Trust, Job security, Sense of belong\\-ing, Equity, Autonomy, Empower\\-ment, Work in successful company,\\ Employee involvement etc.\end{tabular} 
& \begin{tabular}[c]{@{}l@{}} 15 de-motivators - Risk, Stress, \\Inequity, Unfair reward system, \\Poor communication, Poor \\management, Producing poor\\ quality software, Outsourcing Lack\\ of promotion Poor pay, Unrealistic\\ goals, Poor cultural fit, Lack\\ of influence, etc.\end{tabular} \\ \midrule

\begin{tabular}[c]{@{}l@{}} Franca et al. \\\cite{francca2010designing} \end{tabular}
& \begin{tabular}[c]{@{}l@{}}  Motivation in \\SE; identifying \\factors that \\affect motivation\\ at work - software\\ engineers\end{tabular} 
& \begin{tabular}[c]{@{}l@{}} A survey study -\\ 176 software \\engineers in 20 \\software firms in\\ Brazil \\\end{tabular} 
& \begin{tabular}[c]{@{}l@{}} 15 motivators - Work with people,\\ Work life balance, Problem solving,\\ Meaningful products, Work in \\successful company, Creativity,\\ Challenging goals, Empowerment,\\ Tech development, Dev practices, \\Experimentation, Autonomy, \\Broad personal skills, Decision-\\making, Feedback, Participation\\ in entire dev life cycle, Identi\\-fication with task, Career \\development, Changing routine,\\ Rewards and financial incentives \end{tabular} 
& \begin{tabular}[c]{@{}l@{}}  Only considered motivators \\and did not talk about \\de-motivators \end{tabular} \\ \midrule

\begin{tabular}[c]{@{}l@{}} Franca et al. \\\cite{francca2018motivation}\end{tabular}
& \begin{tabular}[c]{@{}l@{}} Motivation in \\SE; multiple case \\studies in four \\software organi\\-zations in Brazil\end{tabular} 
& \begin{tabular}[c]{@{}l@{}} Semi-structured\\ interviews, Diary \\studies and Docu\\-ment analysis
\end{tabular} 
& \begin{tabular}[c]{@{}l@{}} Behavioural - Focus, Engagement\\Care, Commitment, Hard work \\ Interest; work - Pro-activity, \\Interactivity, Mutual help,\\ Productivity \end{tabular} 
& \begin{tabular}[c]{@{}l@{}} Not focused, Carelessness, Laziness\\ Passivity, Social isolation\\ Helpless, Lack of productivity \end{tabular} \\ \midrule

\begin{tabular}[c]{@{}l@{}} Franca et al. \\\cite{francca2014motivated}\end{tabular}
& \begin{tabular}[c]{@{}l@{}}  Motivation in \\SE; a field\\ study with \\software engineers\\ in Brazil \end{tabular} 
& \begin{tabular}[c]{@{}l@{}} A qualitative study\\- semi-structured\\ interviews of 62 \\software engineers - \\Applied thematic\\ analysis \\\end{tabular} 
& \begin{tabular}[c]{@{}l@{}}  8 categories - Engagement (in\\-volved/ hard-working/ curious/\\ proactive), Happiness (excited/\\ good mood, pleased, upbeat),\\ Focused, Collaboration \\(communicative/ \\cooperative, available), \\Professionalism (punctual,\\ responsible), Productivity,\\ Creativity, Stability (calm) \\ Optimism  \end{tabular} 
& \begin{tabular}[c]{@{}l@{}} Uninvolved, Lazy, Indifferent, \\Passive, Bored, Bad mood, Resentful, \\Depressed, Careless, Loose, Reserved \\ Uncooperative, Unavailable, Absent\\ Troublemaker, Irresponsible, \\Unproductive, Un-creative, Pessimistic \end{tabular} \\ \midrule

\begin{tabular}[c]{@{}l@{}}Franca et al. \\\cite{6337859}\end{tabular}
& \begin{tabular}[c]{@{}l@{}}  Motivation in \\SE; focusing\\ on explanatory \\theory on moti\\-vation in SE \\- small company\\ in Brazil \\ \end{tabular} 
& \begin{tabular}[c]{@{}l@{}} A qualitative study\\- semi-structured \\interviews of 10\\ software engineers, \\2 project managers, \\2 directors (all \\below 30 age) \end{tabular} 
& \begin{tabular}[c]{@{}l@{}} Task characteristics - High learn\\-ing High intellectual challenge\\  High autonomy, high self-efficacy\\  Organizational practices - Recog\\-nition, Career progress support, \\Work lifestyle flexibility, Employee\\ participation, Team - collaborative\\ work high technical competence \end{tabular} 
& \begin{tabular}[c]{@{}l@{}} Task characteristics - Low learning \\ Low intellectual challenge, Low\\ self-efficacy, Low autonomy, \\ Low task identity\\ Organizational practices -  Low career \\progression support, Low salaries,\\ Low customer feedback \\High time-pressure, Team -\\ Low technical competency \end{tabular} \\ \midrule

\begin{tabular}[c]{@{}l@{}}Franca et al. \\\cite{FRANCA201479}\end{tabular}
& \begin{tabular}[c]{@{}l@{}}  Motivation in \\SE; cross case\\ analysis of multi\\-ple independent\\ case studies in \\two organizations \\- focusing on \\software engineers \end{tabular} 
& \begin{tabular}[c]{@{}l@{}} A qualitative study\\-  two independent\\ case studies - inter\\-views, diary studies \\ document analysis \\\end{tabular} 
& \begin{tabular}[c]{@{}l@{}} Organizational - Job stability,\\ Career  progression support, \\Feedback, Intellectual  challenge, \\Team cohesion Task - Autonomy,\\ Learning  opportunities, Collabo\\-ration Technical competence,\\ Individual characteristics \end{tabular} 
& \begin{tabular}[c]{@{}l@{}} Poor career progression support \\ Constant goal changes, Poor \\feedback, Political goal \\priorities \end{tabular} \\ \midrule

\begin{tabular}[c]{@{}l@{}} Sach et al.\\ \cite{RN2582}\end{tabular}
& \begin{tabular}[c]{@{}l@{}}  Motivation in \\SE; a workshop\\ on motivation \\with software \\developers, consul\\-tants and techni\\-cal managers \end{tabular} 
& \begin{tabular}[c]{@{}l@{}} A qualitative study\\-  a workshop with\\ 15 developers 3 \\consultants and 5 \\tech. managers \end{tabular} 
& \begin{tabular}[c]{@{}l@{}} Dev- People, Financial, \\Problem solving, Creative, \\Interesting, Challenge Learning,\\ Autonomy, Fear Management,\\ Satisfaction, Programming\\ Appreciation Tech. Managers \\- People Problem solving, Chall\\-enge,  Financial, Enjoyment, \\Habit, Variety, personal Consul\\-tant - People, Developing, \\Challenging, Creative, Interesting\\  Pay, Big picture\end{tabular} 
& \begin{tabular}[c]{@{}l@{}} Only considered motivators \\and did not talk about \\de-motivators \end{tabular} \\ \midrule

\begin{tabular}[c]{@{}l@{}} Sach et al. \\\cite{6092590}\end{tabular}
& \begin{tabular}[c]{@{}l@{}}  Motivation in \\SE; focusing\\ software engi\\-neers' perceptions \\of factors in \\motivation - in one \\organization \end{tabular} 
& \begin{tabular}[c]{@{}l@{}} A qualitative study\\- semi-structured\\ interviews of 13 \\software engineers \\- developers - \\All male \end{tabular} 
& \begin{tabular}[c]{@{}l@{}} Useful work, Producing \\ good software, Learning, Owner\\-ship Solving problems, Experi\\-ence  Collaboration, Exposure, \\Career goals, Company culture, \\People \end{tabular} 
& \begin{tabular}[c]{@{}l@{}} Obstacles, Repetition \\ Uninteresting work, Uncertainty \\ Maintenance, Tedious work \end{tabular} \\ \midrule

\begin{tabular}[c]{@{}l@{}}Tessem \&\\ Maurer\\\cite{tessem2007job}\end{tabular}
& \begin{tabular}[c]{@{}l@{}}  Motivation in \\SE; a case study\\ in a large\\ ICT company - \\Software developers \\from a team \end{tabular} 
& \begin{tabular}[c]{@{}l@{}} A qualitative study\\-  semi-structured \\interviews of 5 practi\\-tioners (3 developers,\\ 1 QA, 1 database \\specialist), general \\observations of the \\team \\\end{tabular} 
& \begin{tabular}[c]{@{}l@{}} Autonomy in work, Variety \\ Significance, Feedback via \\direct communication, Ability\\ to complete whole task \end{tabular} 
& \begin{tabular}[c]{@{}l@{}} Little feedback on tech issues \\ Pressure on work, Feeling \\ of waste of time, Inability to influence \\devs (by QA) \end{tabular} \\ \midrule

\begin{tabular}[c]{@{}l@{}}Khan \&\\ Akbar \\\cite{RN2215} \end{tabular}
& \begin{tabular}[c]{@{}l@{}} Motivation in \\requirements  change \\management\\ (RCM) in GSD\end{tabular} 
& \begin{tabular}[c]{@{}l@{}} An SLR and \\a survey \\study
\end{tabular} 
& \begin{tabular}[c]{@{}l@{}} Extracted 25 motivators-\\ such as accountability, need \\ for change, Requirements trace\\-ability,  RCM task allocation \\across GSD sites, Understanding\\ requested changes, proper task \\coupling, etc.\end{tabular} 
& \begin{tabular}[c]{@{}l@{}} Extracted 25 motivators in \\requirements change management\\ and developed taxonomies of \\identified  motivators.\end{tabular} \\ \midrule

\colorbox{lightgray}{\textbf{This Study}}
&  \colorbox{lightgray}{\textbf{\begin{tabular}[c]{@{}l@{}} Motivation in \\RE;- focusing \\on practitioners\\ who are mainly \\involved in RE\\-related activities \end{tabular} }}
 & \colorbox{lightgray}{\textbf{\begin{tabular}[c]{@{}l@{}} A qualitative \\study - 21 semi-struc\\-tured interviews, \\2 iterations \\ STGT analysis   \end{tabular} }}
& \colorbox{lightgray}{\textbf{\begin{tabular}[c]{@{}l@{}} List of motivating/demoti\\-vating factors along with \\its' effect in RE  - Figure \ref{Theory} \\- Categorised into technical, \\collaborative \& human,  socio-\\technical and social. Collabo\\-rative \& human has three sub-\\categories - individual,  team \\and customer related. Identi\\-fied the impact of these factors\\ on RE\end{tabular} }}
& \colorbox{lightgray}{\textbf{\begin{tabular}[c]{@{}l@{}} Identified demotivating factors \\ for each of above categories -\\ (e.g. Tech - incomplete requirements,\\ Individual - personal issues, Team - \\team conflicts, Customers - difficult \\to handle cust., Socio-tech - domain\\ knowledge, Social - gender biases\end{tabular}}} \\ \midrule
\end{tabular}%
}
\end{table*}

\subsection{\textbf{Comparison with existing studies} }  \label{5.3}
In our study, we identified a set of motivating/demotivating factors, which we categorised as ``technical, collaborative \& human, socio-technical and social" based on data analysis we conducted. Referring to prior studies, a variety of studies identified a set of (de)motivating factors in SE. 
As shown in the table \ref{Table: comparison with related work}, almost all the studies main focus was on software developers and were not talking about RE-related activities except in \cite{RN2215}, which was only focusing requirements change management in GSD context. Further, the identified motivating/demotivating factors were mostly related to individual, organization, team aspects which covers similar set of factors. For example, autonomy, feedback, collaboration with others, work-life balance, challenging/interesting work, rewards \& incentives were identified in the majority of studies. The similar factors were identified in our study and categorised under ``collaborative \& human", but in-detail along with the impact of each (de)motivating factor on RE-related activities and/or SE in general, which is missing in most of the studies. The majority of these related work were mainly limited to software engineers in one or two organizations in a particular country (\cite{francca2018motivation}, \cite{francca2014motivated}, \cite{6337859}, \cite{FRANCA201479}) which can be a limitation in identifying (de)motivating factors in the broader SE context. Among these studies, a few only focused on identifying motivating factors without considering (de)motivating factors where it only discussed the positive aspects \cite{francca2010designing} \cite{RN2582}. The majority of these studies were built on the work carried out by  Beecham et al. \cite{BEECHAM2008860} and sharp et al. \cite{sharp2009models}, where they highlighted the need of building a rich descriptive and context sensitive phenomenon of motivation focusing on diverse contexts, and roles in the SE context, which we tried to address in the area of requirements engineering, which has been a limited focus related to motivation. 

We identified that several motivational theories from psychology research  have been taken into consideration. These theories tend to focus on the factors that satisfy human needs or dynamic processes through which motivation occurs \cite{francca2013motivation} \cite{sharp2009models}. Based on these, the motivation theories are described as content theories, such as Maslow's Hierarchy of Needs \cite{maslow1987motivation}, Herzberg's Motivation-Hygiene theory \cite{herzberg2017motivation}, McClelland’s Needs theory \cite{mcclelland1961achieving} and process theories, such as Vroom’s Expectancy theory \cite{RN3001}, Hackman and Oldham’s Job Characteristics theory \cite{hackman1976motivation}, and Locke’s Goal-setting theory \cite{locke1968toward}. However, the systematic review conducted by Hall et al., \cite{hall2009systematic} highlighted that these classic motivation theories do not explicitly interpret software engineers' motivation. They identified this by analysing 92 studies of motivation in SE on the use of motivation theories to investigate the effect of motivation in SE. Further, they emphasised the need for more empirical studies to develop a strong body of knowledge on motivation in software engineering rather than limiting it to existing theories. In another study, Sharp et al. \cite{sharp2009models} also pointed out that these theories are mainly applicable to the management perspective in the SE context as none of them is context specified and tends to cover general management aspects of motivation. Further, based on the literature, they proposed a motivation model (MOCC model) with four key components; \emph{motivators}, \emph{outcomes}, \emph{characteristics} and \emph{context} in the SE context. However, they emphasised the importance of evaluating the proposed model and the need to have a definitive model of motivation that can be used adequately to capture motivators and demotivators in the SE context. Moreover, systematic and empirical studies conducted by Beecham et al. \cite{BEECHAM2008860}, and Franca et al.  \cite{francca2010designing} \cite{francca2011motivation} \cite{francca2013motivation} also highlighted that as motivation is context-dependent, necessary theories and models need to be developed to explore software practitioners motivation to perform their allocated tasks in the SE context. A detailed analysis of the use of these motivation theories and to what extent they have been used in the RE context is presented in the SLR we conducted \cite{RN1600}.  

\subsection{\textbf{Implications}} \label{section 5.4}
Our findings suggest several implications for both researchers and software practitioners who are involved in RE-related activities. This section discusses these implications and corresponding ideas for future work. 

\subsection{Implications for Researchers} The main contribution of this research is \textbf{the theory } (figure \ref{Theory}) about the influence of motivation on RE developed using STGT approach via 21 in-depth interviews. 
This framework can be used to understand better how software practitioners' motivation influences RE-related activities. In future work, researchers can work on the validation of the theory we developed in similar contexts and also can expand this research by exploring it with various other human aspects in different settings (e.g., in different contexts). Our theory is developed using STGT approach \cite{RN1609} and presented using the well-defined Strauss \& Corbin's coding paradigm, including context, causal conditions, intervening conditions, strategies and consequences 
to better understand the influence of motivation on RE-related activities \cite{strauss1990basics}. Hence, understanding and applying this theory framework in other contexts, such as replicating it for a specific region, culture, or organization, would be easier and based on the findings, the researchers can identify whether the software practitioners use the same strategies irrespective of their contextual differences. Further, a comparison study can be conducted to identify similarities and differences of motivating/demotivating factors that might impact diverse SE activities such as RE, coding, and testing, which will be helpful in identifying unique motivating/demotivating factors for SE activities.

\subsection{Implications for Software Practitioners}
This research has implications for software practitioners involved in RE-related activities in agile software development projects. Our findings can be used as \textbf{a guide for practitioners, specifically managers and team leads} to help understand and manage the influence of motivation on RE-related activities. Software practitioners who are involved in RE-related activities can benefit from these findings in multiple ways. 
\par Software practitioners can gain a \textbf{better understanding of types of motivation factors and their impacts} from the list of motivating and demotivating factors we identified as intervening conditions. With that, they could tackle the constraints of demotivating factors that can be aroused during RE-related activities. Further, the managers/ team leads can pay attention to identifying motivating/demotivating factors that can influence RE-related activities and play a supporting role in keeping up the motivating situations or overcoming demotivating situations in the team. They also can focus on \textbf{using the identified strategies to manage the motivation of their teams} in similar contexts. For example, mapping between the identified motivating/demotivating factors and the strategies identified can be used to find the most suitable strategies to overcome the negative influences of demotivating factors in their contexts. For instance, conflicts among the team or with customers are considered as a demotivating factor according to the majority of the participants, resulting in missing/ incomplete requirements, and to overcome this situation, strategies such as rebuild collaboration (S2), open \& individual discussions (S3) and changing the process, or people involved (S7) can be used by the managers or team leads. The consequences we identified by applying these strategies can also be helpful in understanding the impact of motivation on RE-related activities and the overall software development process. 
\par With these strategies and the consequences we identified from our analysis, the following are some of our recommendations on reinforcing motivated situations or mitigating demotivated situations for software practitioners when they are involved in RE-related activities. 
\par \faHandORight\hspace{0.1cm} \textbf{More exposure to experience \& learn by providing diverse, challenging tasks:} As every participant highlighted that they get motivated when they are given the opportunity to explore, experiment and learn with their allocated tasks, the managers/ team leads can pay more attention to allocate diverse, challenging tasks so that the practitioners can use various approaches to complete their tasks. This can be helpful in conducting quality RE-related activities from the early stages of the project (e.g., using various relevant techniques to elicit requirements) - S1. 

\par \faHandORight\hspace{0.1cm} \textbf{Open or individual discussions, collaborations and regular updates to avoid conflicts:} Interactions/ collaborations/ communication among team members and with customers   play a major part in motivating software practitioners and avoiding conflicts. Hence, it is important for managers to initiate open discussions with both team members and customers, individual discussions when required to enhance collaborations and keep everyone (team members \& customers) updated, especially in prioritising and managing requirements - S2, S3, S9.

\par \faHandORight\hspace{0.1cm} \textbf{Being empathetic towards team members and customers/ clients/ end users:} Demotivating situations are unavoidable in any project as the software development process mainly depends on people. Practitioners involved in RE-related activities have to interact with a different set of people, such as clients/ end-users who may not have an IT background, as well as technical people, such as developers/testers. Therefore, it is important to be more empathetic towards both of these categories as they have to deal with their emotions, personalities, and behaviours when doing RE-related activities. By trying to be more empathetic towards other people, practitioners can minimise the demotivating situations as it is helpful in better understanding both ends (developers \& clients/end users) - S6. 

\par \faHandORight\hspace{0.1cm} \textbf{Make decisions according to the situation:} Our findings suggest that both managers/ team leads and team members should have the ability to make decisions when required in accordance with the situations to overcome demotivating situations and enhance the motivation of the team. For example, managers/team leads should take the necessary actions, such as changing the process of doing RE-related activities, changing the people involved in those activities by re-allocating the tasks or seeking senior management support before making the situation more complicated with demotivated team members and not getting expected quality outcome from their allocated tasks. Further, team members should also have the ability to directly discuss their issues (e.g., in sprint retrospectives) that can demotivate them and sometimes move out from the demotivated situations if things are unavoidable, unable to solve or not being addressed by the management - S7, S8, S10, S12. 

\section{\textbf{Threats to Validity}} \label{section 6}
\textbf{Threats to external validity: } 
Our findings are limited to the agile software development context, with 19 of 21 participants using agile methods. Although two participants (INT03 \& INT09) occasionally used the waterfall method, their examples primarily involved agile projects. The data collection lacks global representation, predominantly featuring participants from Australia. Our findings are most relevant to the participants' experiences, their organizations' practices, and their country of residence, potentially extending to similar contexts but not generalizing to the global RE/SE community. However, in practice, such generalization is unlikely to be achievable \cite{masood2020agile}. 
In our interviews, we focused on asking in-depth questions about participants' involvement in RE-related activities and their perspectives and experience on the impact of motivation when involved in RE-related activities. The participants have various experiences related to various domains where their involvement in RE may vary with the context. Hence, the theory we developed can be used in the defined contextual and causal conditions (section \ref{section 4.2} \& \ref{section 4.3}), we suggest conducting more empirical case studies expanding our theory focusing on different software organizations and domains. 

\par  \textbf{Threats to internal validity: } 
There could be misinterpretations/ misunderstandings between what we asked and our participants' responses to it. To mitigate this, we used follow-up questions during the interviews and got further clarifications on their statements. All of the authors contributed to developing the interview guide. Although the main coding was done by the first author, all the codes, concepts and categories were collaboratively discussed and finalized by all the authors. Further, by providing a number of interview quotes as examples, we mitigate any reporting biases in this study.

\section{\textbf{Conclusion}} \label{section 8}
In this study, we demonstrated how the motivation of software practitioners influences RE-related activities in agile software development projects. We conducted a socio-technical grounded theory study (STGT) \cite{RN1609}, by conducing 21 in-depth interviews of software practitioners who are predominantly involved in RE-related activities. We developed a theory of understanding the influence of motivation on RE-related activities using the full STGT method.
This theory explains how the motivation of software practitioners impacts RE-related activities and SE in general within the context such as country and project domains that they are working on, causal conditions such as practitioners' involvement in RE-related activities when working on the agile software development projects and a set of intervening conditions that are identified as motivating/demotivating factors of software practitioners which can give rise to or mediate the influence of motivation on RE-related activities. We also identified a set of strategies that can be applied by software practitioners to keep up the motivating situations or mitigate the demotivating situations when involved in RE-related activities, resulting in a set of consequences. From these findings, we also provide several implications for both the research community and software practitioners, along with specific recommendations that can be used by managers/ team leads to better understand and work on mitigating the negative influences that have occurred due to demotivating situations. 

\begin{acknowledgements}
This work is supported by Monash Faculty of IT PhD scholarships. Hidellaarachchi and Grundy are supported by ARC Laureate Fellowship FL190100035 and this work is also partially supported by ARC Discovery Project DP200100020.
\end{acknowledgements}

%
\section*{Declarations}
\textbf{Conflict of Interests:} The authors declare that they have no conflict of interest.\\
\textbf{Data Availability Statement:} All data generated or analysed during this study are included in this published article (and its supplementary information files).

\bibliographystyle{spmpsci}      
\bibliography{output.bib}

\begin{thebibliography}{10}
\providecommand{\url}[1]{{#1}}
\providecommand{\urlprefix}{URL }
\expandafter\ifx\csname urlstyle\endcsname\relax
  \providecommand{\doi}[1]{DOI~\discretionary{}{}{}#1}\else
  \providecommand{\doi}{DOI~\discretionary{}{}{}\begingroup \urlstyle{rm}\Url}\fi

\bibitem{ali2019effective}
Ali, Z., Yaseen, M., Ahmed, S.: Effective communication as critical success factor during requirement elicitation in global software development.
\newblock International Journal of Computer Science Engineering (IJCSE) \textbf{8}(3), 108--115 (2019)

\bibitem{BEECHAM2008860}
Beecham, S., Baddoo, N., Hall, T., Robinson, H., Sharp, H.: Motivation in software engineering: A systematic literature review.
\newblock Information and Software Technology \textbf{50}(9), 860--878 (2008).
\newblock \doi{https://doi.org/10.1016/j.infsof.2007.09.004}.
\newblock \urlprefix\url{https://www.sciencedirect.com/science/article/pii/S0950584907001097}

\bibitem{RN2731}
Chakraborty, A., Baowaly, M., A~Arefín, U., Bahar, A.N.: The role of requirement engineering in software development life cycle.
\newblock Journal of Emerging Trends in Computing and Information Sciences \textbf{3}, 723--729 (2012)

\bibitem{RN2713}
Deak, A., Stålhane, T., Sindre, G.: Challenges and strategies for motivating software testing personnel, vol.~73.
\newblock Butterworth-Heinemann (2016).
\newblock \doi{10.1016/j.infsof.2016.01.002}.
\newblock \urlprefix\url{https://doi-org.ezproxy.lib.monash.edu.au/10.1016/j.infsof.2016.01.002}

\bibitem{demarco2013peopleware}
DeMarco, T., Lister, T.: Peopleware: productive projects and teams.
\newblock Addison-Wesley (2013)

\bibitem{francca2013motivation}
Fran{\c{c}}a, A.C.C., de~Ara{\'u}jo, A.C., Da~Silva, F.Q.: Motivation of software engineers: A qualitative case study of a research and development organisation.
\newblock In: 2013 6th International Workshop on Cooperative and Human Aspects of Software Engineering (CHASE), pp. 9--16. IEEE (2013)

\bibitem{francca2011motivation}
Fran{\c{c}}a, A.C.C., Gouveia, T.B., Santos, P.C., Santana, C.A., da~Silva, F.Q.: Motivation in software engineering: A systematic review update.
\newblock In: 15th Annual Conference on Evaluation \& Assessment in Software Engineering (EASE 2011), pp. 154--163. IET (2011)

\bibitem{francca2010designing}
Fran{\c{c}}a, A.C.C., da~Silva, F.Q.: Designing motivation strategies for software engineering teams: an empirical study.
\newblock In: Proceedings of the 2010 ICSE Workshop on Cooperative and Human Aspects of Software Engineering, pp. 84--91 (2010)

\bibitem{francca2018motivation}
Fran{\c{c}}a, C., Da~Silva, F.Q., Sharp, H.: Motivation and satisfaction of software engineers.
\newblock IEEE Transactions on Software Engineering \textbf{46}(2), 118--140 (2018)

\bibitem{francca2014motivated}
Fran{\c{c}}a, C., Sharp, H., Da~Silva, F.Q.: Motivated software engineers are engaged and focused, while satisfied ones are happy.
\newblock In: Proceedings of the 8th ACM/IEEE International Symposium on Empirical software Engineering and Measurement, pp. 1--8 (2014)

\bibitem{6337859}
França, A.C.C., Carneiro, D.E.S., Silva, F.Q.B.d.: Towards an explanatory theory of motivation in software engineering: A qualitative case study of a small software company.
\newblock In: 2012 26th Brazilian Symposium on Software Engineering, pp. 61--70 (2012).
\newblock \doi{10.1109/SBES.2012.28}

\bibitem{FRANCA201479}
França, A.C.C., {da Silva}, F.Q., de~L.C.~Felix, A., Carneiro, D.E.: Motivation in software engineering industrial practice: A cross-case analysis of two software organisations.
\newblock Information and Software Technology \textbf{56}(1), 79--101 (2014).
\newblock \doi{https://doi.org/10.1016/j.infsof.2013.06.006}.
\newblock \urlprefix\url{https://www.sciencedirect.com/science/article/pii/S0950584913001353}.
\newblock Special sections on International Conference on Global Software Engineering – August 2011 and Evaluation and Assessment in Software Engineering – April 2012

\bibitem{garden1988behavioural}
Garden, A.M.: Behavioural and organisational factors involved in the turnover of high tech professionals.
\newblock ACM SIGCPR Computer Personnel \textbf{11}(4), 6--9 (1988)

\bibitem{RN3000}
Golembiewski, R.T.: Handbook of organizational behavior, revised and expanded.
\newblock CRC Press (2000)

\bibitem{hackman1976motivation}
Hackman, J.R., Oldham, G.R.: Motivation through the design of work: Test of a theory.
\newblock Organizational behavior and human performance \textbf{16}(2), 250--279 (1976)

\bibitem{RN2572}
Hall, T., Baddoo, N., Beecham, S., Robinson, H., Sharp, H.: A systematic review of theory use in studies investigating the motivations of software engineers, vol.~18.
\newblock Association for Computing Machinery (2009).
\newblock \doi{10.1145/1525880.1525883}.
\newblock \urlprefix\url{https://doi-org.ezproxy.lib.monash.edu.au/10.1145/1525880.1525883}

\bibitem{hall2009systematic}
Hall, T., Baddoo, N., Beecham, S., Robinson, H., Sharp, H.: A systematic review of theory use in studies investigating the motivations of software engineers.
\newblock ACM Transactions on Software Engineering and Methodology (TOSEM) \textbf{18}(3), 1--29 (2009)

\bibitem{hertel2003motivation}
Hertel, G., Niedner, S., Herrmann, S.: Motivation of software developers in open source projects: an internet-based survey of contributors to the linux kernel.
\newblock Research policy \textbf{32}(7), 1159--1177 (2003)

\bibitem{herzberg2017motivation}
Herzberg, F.: Motivation to work.
\newblock Routledge (2017)

\bibitem{RN1600}
Hidellaarachchi, D., Grundy, J., Hoda, R., Madampe, K.: The effects of human aspects on the requirements engineering process: A systematic literature review.
\newblock IEEE Transactions on Software Engineering  (2021).
\newblock \doi{10.1109/TSE.2021.3051898}

\bibitem{10.1145/3546943}
Hidellaarachchi, D., Grundy, J., Hoda, R., Mueller, I.: The influence of human aspects on requirements engineering-related activities: Software practitioners’ perspective.
\newblock ACM Trans. Softw. Eng. Methodol.  (2022).
\newblock \doi{10.1145/3546943}.
\newblock \urlprefix\url{https://doi.org/10.1145/3546943}

\bibitem{RN1609}
Hoda, R.: Socio-technical grounded theory for software engineering.
\newblock IEEE Transactions on Software Engineering  (2021).
\newblock \doi{10.1109/TSE.2021.3106280}

\bibitem{7320433}
Johann, T., Maalej, W.: Democratic mass participation of users in requirements engineering?
\newblock In: 2015 IEEE 23rd International Requirements Engineering Conference (RE), pp. 256--261 (2015).
\newblock \doi{10.1109/RE.2015.7320433}

\bibitem{john2005human}
John, M., Maurer, F., Tessem, B.: Human and social factors of software engineering: workshop summary.
\newblock ACM SIGSOFT Software Engineering Notes \textbf{30}(4), 1--6 (2005)

\bibitem{RN2997}
Johnson, J.A.: Measuring thirty facets of the five factor model with a 120-item public domain inventory: Development of the ipip-neo-120.
\newblock Journal of Research in Personality \textbf{51}, 78--89 (2014).
\newblock \doi{https://doi.org/10.1016/j.jrp.2014.05.003}.
\newblock \urlprefix\url{http://www.sciencedirect.com/science/article/pii/S0092656614000506}

\bibitem{jones1991applied}
Jones, C.: Applied software measurement: assuring productivity and quality.
\newblock McGraw-Hill, Inc. (1991)

\bibitem{RN2215}
Khan, A.A., Akbar, M.A.: Systematic literature review and empirical investigation of motivators for requirements change management process in global software development.
\newblock Journal of Software: Evolution and Process \textbf{n/a}(n/a), e2242.
\newblock \doi{10.1002/smr.2242}.
\newblock \urlprefix\url{https://onlinelibrary.wiley.com/doi/abs/10.1002/smr.2242}

\bibitem{kolpondinos2017tailoring}
Kolpondinos, M.Z.H., Glinz, M.: Tailoring gamification to requirements elicitation: A stakeholder-centric motivation concept.
\newblock In: 2017 IEEE/ACM 10th International Workshop on Cooperative and Human Aspects of Software Engineering (CHASE), pp. 9--15. IEEE (2017)

\bibitem{locke1968toward}
Locke, E.A.: Toward a theory of task motivation and incentives.
\newblock Organizational behavior and human performance \textbf{3}(2), 157--189 (1968)

\bibitem{madampe2022role}
Madampe, K., Hoda, R., Grundy, J.: The role of emotional intelligence in handling requirements changes in software engineering.
\newblock arXiv preprint arXiv:2206.11603  (2022)

\bibitem{10061282}
Madampe, K., Hoda, R., Grundy, J.: A framework for emotion-oriented requirements change handling in agile software engineering.
\newblock IEEE Transactions on Software Engineering pp. 1--20 (2023).
\newblock \doi{10.1109/TSE.2023.3253145}

\bibitem{maslow1987motivation}
Maslow, A.H., Frager, R., Fadiman, J., McReynolds, C., Cox, R.: Motivation and personality (3rd).
\newblock New York  (1987)

\bibitem{masood2020agile}
Masood, Z., Hoda, R., Blincoe, K.: How agile teams make self-assignment work: a grounded theory study.
\newblock Empirical Software Engineering \textbf{25}(6), 4962--5005 (2020)

\bibitem{mcclelland1961achieving}
McClelland, D.C., Mac~Clelland, D.C.: Achieving society, vol. 92051.
\newblock Simon and Schuster (1961)

\bibitem{murphy2010human}
Murphy, G.C.: Human-centric software engineering.
\newblock In: Proceedings of the FSE/SDP workshop on Future of software engineering research, pp. 251--254 (2010)

\bibitem{PROCACCINO2005194}
Procaccino, J.D., Verner, J.M., Shelfer, K.M., Gefen, D.: What do software practitioners really think about project success: an exploratory study.
\newblock Journal of Systems and Software \textbf{78}(2), 194--203 (2005).
\newblock \doi{https://doi.org/10.1016/j.jss.2004.12.011}.
\newblock \urlprefix\url{https://www.sciencedirect.com/science/article/pii/S0164121204002614}

\bibitem{rasch1992factors}
Rasch, R.H., Tosi, H.L.: Factors affecting software developers' performance: An integrated approach.
\newblock MIS quarterly pp. 395--413 (1992)

\bibitem{richens1998hr}
Richens, E.: Hr strategies for is professionals in the 21st century.
\newblock In: Proceedings of the 1998 ACM SIGCPR conference on computer personnel research, pp. 289--291 (1998)

\bibitem{RN3002}
Robbins, S.P., Judge, T.: Essentials of organizational behavior, vol.~7.
\newblock Prentice Hall Upper Saddle River, NJ (2003)

\bibitem{RN2582}
Sach, R., Sharp, H., Petre, M.: Continued involvement in software development: motivational factors.
\newblock Proceedings of the 2010 ACM-IEEE International Symposium on Empirical Software Engineering and Measurement. Association for Computing Machinery, Bolzano-Bozen, Italy (2010).
\newblock \doi{10.1145/1852786.1852843}.
\newblock \urlprefix\url{https://doi-org.ezproxy.lib.monash.edu.au/10.1145/1852786.1852843}

\bibitem{6092590}
Sach, R., Sharp, H., Petre, M.: Software engineers' perceptions of factors in motivation: The work, people, obstacles.
\newblock In: 2011 International Symposium on Empirical Software Engineering and Measurement, pp. 368--371 (2011).
\newblock \doi{10.1109/ESEM.2011.50}

\bibitem{sharp2009models}
Sharp, H., Baddoo, N., Beecham, S., Hall, T., Robinson, H.: Models of motivation in software engineering.
\newblock Information and software technology \textbf{51}(1), 219--233 (2009)

\bibitem{snijders2015refine}
Snijders, R., Dalpiaz, F., Brinkkemper, S., Hosseini, M., Ali, R., Ozum, A.: Refine: A gamified platform for participatory requirements engineering.
\newblock In: 2015 IEEE 1st International Workshop on Crowd-Based Requirements Engineering (CrowdRE), pp. 1--6. IEEE (2015)

\bibitem{strauss1990basics}
Strauss, A., Corbin, J.: Basics of qualitative research.
\newblock Sage publications (1990)

\bibitem{sutcliffe2002user}
Sutcliffe, A.: User-centred requirements engineering.
\newblock Springer Science \& Business Media (2002)

\bibitem{tarawneh2011suggested}
Tarawneh, H., et~al.: A suggested theoretical framework for software project success.
\newblock Journal of Software Engineering and Applications \textbf{4}(11), 646 (2011)

\bibitem{tessem2007job}
Tessem, B., Maurer, F.: Job satisfaction and motivation in a large agile team.
\newblock In: International Conference on Extreme Programming and Agile Processes in Software Engineering, pp. 54--61. Springer (2007)

\bibitem{4685653}
Thew, S., Sutcliffe, A.: Investigating the role of 'soft issues' in the re process.
\newblock In: 2008 16th IEEE International Requirements Engineering Conference, pp. 63--66 (2008).
\newblock \doi{10.1109/RE.2008.35}

\bibitem{RN3001}
Vroom, V.H.: Work and motivation  (1964)

\end{thebibliography}
%
%

\appendix
\section{Appendix : Interview Protocol} \label{A}
\begin{footnotesize}
\textbf{Section 01: Demographic Information} 
\begin{itemize}
    \item Can you briefly tell me about yourself? 
    \item Considering your experience; how many years of experience do you have in the software industry?
    \item How many years of experience do you have in requirement engineering these years?
    \item Can you think of a past or a current software engineering project where you were engaged in RE-related activities? 
    \begin{itemize}
        \item Can you please tell me very briefly about that project?
        \item Were you developing a product or service?
        \item What software development methods are/were you using?
        \item What are the size and composition of the team?
        \item What is/was your role on the project?
        \item What RE-related activities were being performed on that project? 
        \item What was your involvement in those activities?
    \end{itemize}
\end{itemize}
\textbf{Section 02: Views on the influence of motivation 0n RE-related activities} 
\par \textbf{Motivation} can be described as the willingness to do a certain action, for example, some people are motivated to work when they are given good feedback. \\
Talking about motivation;

\begin{itemize}
    \item In your opinion, what are the aspects that motivate you when involved in RE-related activities?
    \item Can you explain how these aspects motivate you to perform RE-related activities? (Ask for more details about the factors they mentioned)
    \item Can you share your experience on how these motivation aspects impact on how you approach requirements engineering activities?
      \item Can you think of the most motivated team that you have ever been part of when involved in RE-related activities? 
    \begin{itemize}
        \item Can you tell me why you think it is a highly motivated team?
        \item Did you feel highly motivated in that team? If so, what are the reasons for you to be highly motivated in that team?
    \end{itemize}
    \item Now referring to the opposite of what you have mentioned, in your opinion, what are the aspects that demotivate you when involved in RE-related activities?
    \item Can you share your experience on how these demotivating factors impact on how you approach requirements engineering activities?
     \item Can you think of the most demotivated team (if any) that you have ever been part of when involved in RE-related activities? 
    \begin{itemize}
        \item Can you tell me why you think it is the demotivated team?
        \item How did you handle the demotivating situation to get your work done?
        \item Does your manager/ team leads/ senior management use any strategies to overcome demotivated situations? 
        \item If so, can you explain what are those?
        \item Can you elaborate on how the demotivated situation impacted you to complete your tasks?
        \item Did you attempt to stay motivated and make the project completed? If so, can you explain how?
    \end{itemize}
    \item Any final thoughts about the impact of motivation on RE-related activities?
\end{itemize}
\end{footnotesize}

\section{Appendix B: Detailed List of Collaborative \& Human Motivation Factors and their influence on RE} \label{B}

\begin{itemize}
    \item Table \ref{Table : impact of human-related motivation factors}: Influence of collaborative \& human motivation factors on performing RE-related activities (Individual-related)
    \item Table \ref{Table : impact of human-related motivation factors -Team}: Influence of collaborative \& human motivation factors on performing RE-related activities (Team-related)
    \item Table \ref{Table : impact of human-related motivation factors-customers}: Influence of collaborative \& human motivation factors on performing RE-related activities (Customer/Clients/End user-related)
\end{itemize}

\begin{table*}[]
\caption{Influence of collaborative \& human motivation factors on performing RE-related activities (Individual-related)}
\label{Table : impact of human-related motivation factors}
\resizebox{\linewidth}{!}{%
\begin{tabular}{@{}clll@{}}
\toprule
\textbf{Sub-Category}                                                                   & \textbf{\begin{tabular}[c]{@{}l@{}}Motivating/ \\ demotivating Factors\end{tabular}}                                            & \textbf{Impact on RE/SE} & \textbf{Responses} \\ \midrule
\multirow{15}{*}{\begin{tabular}[c]{@{}l@{}} Individual\end{tabular}}                                                      & (+) Job satisfaction                                               & \begin{tabular}[c]{@{}l@{}} Feeling of doing something useful leads to collect all the\\ requirements needed\\ Leads to happiness \& satisfaction which motivates to do \\new activities\\ Valuing the work done improves the quality of the solution\\ Satisfied outcomes increase the involvement of other projects \\ Less complains on requirements changes\\ Complete the tasks successfully \end{tabular}  &  21/21 \\  \cmidrule(l){2-4} 
       & (+) Interacting with others                                                    & \begin{tabular}[c]{@{}l@{}} Improve collaborative thinking, helpful in providing\\ advanced solutions \\ Ability to identify missing/ incorrect requirements \\ Enjoy direct conversations with customers to clarify \\unclear requirements \\ Helpful in having in-depth discussions to elicit detailed \\requirements \\ Interacting with the development team to come up with\\ the technically most feasible approach for the client \\requirements\end{tabular}  &  19/21                                                                                      \\ \cmidrule(l){2-4}   
    
    & (+) Individual interest                                          & \begin{tabular}[c]{@{}l@{}} Highly involved in each tasks/activities from the beginning \\Helps to achieve goals \& increase the quality of the service \\ Makes the practitioners create novel/ highly practical features \\ Dealing with challenging customers/situations to collect\\ requirements \end{tabular} & 19/21 \\ \cmidrule(l){2-4} 
    
    & \begin{tabular}[c]{@{}l@{}} (+) Personal satisfaction  \end{tabular}                                      & \begin{tabular}[c]{@{}l@{}}  Salary, incentives, promotions, recognition motivate individuals to\\ achieve any goal\\
The self-satisfaction of doing something useful will make\\ practitioners perform well on given tasks\\ Getting clarity of the role makes practitioners complete \\assigned tasks on time\\Wanting to do something big motivates them to dig deeper\\ into requirements\\
Positive appreciation may helpful in further development   \end{tabular} & 19/21                                                                                                      \\ \cmidrule(l){2-4} 
        & (+)  Willingness to explore                             
        & \begin{tabular}[c]{@{}l@{}} Find efficient ways to gather requirements \\ Learn \& utilise new skills throughout the project \\ Ability to react fast to the system/requirements changes due \\to keeping up with new technologies  \\ Provide advanced designs/ solutions to the problems \\ Enhance individual performance result in quality outcome  \end{tabular}  & 18/21                                              \\ \cmidrule(l){2-4} 
             & (+) Work-life balance                                
             & \begin{tabular}[c]{@{}l@{}} Ability to concentrate on the work properly \\ Improve the quality of the work \\ Less external impact on the tasks \\ Improve personal satisfaction, result in on-time product delivery \end{tabular}      & 17/21     \\\cmidrule(l){2-4} 
        & (+) Career development                                    
        & \begin{tabular}[c]{@{}l@{}} Practitioners' involvement with clients to gain experience for \\career development,result in getting complete requirements \\ Feeling of having personal development by working with the\\ team, increases the engagement with the team, resulting \\in coming up with best solutions\\ Getting experience for long-term career development makes\\ the practitioners learn new skills \& techniques and utilise \\them in RE   \end{tabular}     & 15/21                       \\ \cmidrule(l){2-4} 
                & (+) Hard-working nature                                                         & \begin{tabular}[c]{@{}l@{}} Doing work beyond expectations and  completing tasks\\ before the deadlines \\ Keep on trying to improve, resulting in improving\\ the quality of the work \\ Getting detailed requirements from difficult to\\ handle customers \\Committed to doing the work, resulting in successful\\ project completion \end{tabular}  & 15/21                                                \\ \cmidrule(l){2-4} 
         
         & (+) Decision-making power                                                 
         & \begin{tabular}[c]{@{}l@{}} Easily conduct RE-related activities, specially elicitation \\with customers \\ Ability to take the right decision on time, resulting in\\ no project delays \\ Decide whom to approach to have a direct talk/ \\long discussions on clients' side \\ Helpful to be more responsible and giving 100\% each task \end{tabular}                                                                                   & 12/21     \\\cmidrule(l){2-4}                     
         & (+) Empathize with others                                                         & \begin{tabular}[c]{@{}l@{}} Helpful in understanding customers and their requirements \\ Ability to design the system with less confusion \\ Successfully play the intermediary role between \\the client and product development team until on-time product\\ delivery  \end{tabular} & 12/21                                                          \\\cmidrule(l){2-4}
         
         & (+) Problem-solving ability                                  
         & \begin{tabular}[c]{@{}l@{}}   Easily resolve complex requirements \\ Finding the best/creative solution to the problems \\ Understanding exact problem, resulting in getting clear \\requirements \\ High quality final outcome result in customer satisfaction\end{tabular} & 09/21 \\\cmidrule(l){2-4}
         
         & (+) Negotiation skills                                                                     & \begin{tabular}[c]{@{}l@{}} Helpful in prioritizing requirements \\ Important in dealing with customers on requirements changes \\  Negotiations among team \& customers on using\\ different approaches to achieve specific goals \end{tabular}    & 7/21                                                             \\\cmidrule(l){2-4}
         
        & (+/-) Language \& communication                                         
        & \begin{tabular}[c]{@{}l@{}}  Good communication helps to obtain clear, detailed \\requirements \\ Make the work efficient by direct, effective communication \\ Obtain clear descriptions of what the system supposed\\ to do which is essential for good performance \\ Language barriers create difficulties on understanding the problem \\ Taking more time to complete tasks due to poor communication \end{tabular}  & 10/21                                                  \\\cmidrule(l){2-4}
         &     (-) Personal issues                                                            & \begin{tabular}[c]{@{}l@{}} Personal issues related to families creates less involvement, resulting\\ in incomplete tasks \\ Insecure feelings at work create conflicts with others, resulting\\ in quality of the work \\ \end{tabular}    & 7/21           \\\cmidrule(l){2-4}
& (-) Irresponsible behaviour                                    
& \begin{tabular}[c]{@{}l@{}} Wasting time on completing tasks, result in exceeding the timeline \& \\budget of the project \\ Create conflicts between the team and the clients, resulting in getting \\incomplete requirements \\ Completing tasks just to get by, impacting the quality of the final\\ outcome of the project\end{tabular} & 6/21 

\\ \midrule
\end{tabular}%
}
\end{table*}

 \begin{table*}[]
\caption{Influence of collaborative \& human motivation factors on performing RE-related activities (Team-related)}
\label{Table : impact of human-related motivation factors -Team}
\resizebox{\linewidth}{!}{%
\begin{tabular}{@{}clll@{}}
\toprule
\textbf{Sub-Category}                                                                   & \textbf{\begin{tabular}[c]{@{}l@{}}Motivating/ \\ demotivating Factors\end{tabular}}                                            & \textbf{Impact on RE/SE} & \textbf{Responses} \\ \midrule

\multirow{16}{*}{\begin{tabular}[c]{@{}l@{}} Team \end{tabular}} & \begin{tabular}[c]{@{}l@{}} (+) Cooperative team members \end{tabular}                                                      & \begin{tabular}[c]{@{}l@{}} Support \& opinions of the team easily solve the problems\\ by making decisions and solutions together \\  Cooperation among the team will increase the ability \\to work well by helping each other, result in final\\ product in line with customer expectations \\
Friendly collaboration will make the process fun \\and effective \\
Collaboration  with the team is helpful in knowledge \\sharing, doing RE clearly,\\ making the product delivery on-time\\ 
Helps to come up with better solutions with the \\collaboration of diverse people \\ Helpful in understanding complex requirements \end{tabular}  & 20/21                                                                                                                
     \\ \cmidrule(l){2-4} 


& \begin{tabular}[c]{@{}l@{}} (+) Knowledgeable team\\ members   \end{tabular}                                            
& \begin{tabular}[c]{@{}l@{}} Understand complex requirements in particular domains \\ Helpful in negotiating technical feasibility with customers \\ Ability to solve problems easily \\ Adding value to the project with their knowledge \& skills, \\result in quality outcome \\ Make it easy to deal with cross-functional teams in\\ requirements gathering \end{tabular}  &  19/21 \\  \cmidrule(l){2-4} 
       
       & (+) Passionate team members                                                    
       & \begin{tabular}[c]{@{}l@{}} Provide high-quality outcome/solution \\  Willing to learn new skills \&  keep updated with technology, \\benefited in achieving desired outcomes \\ Focus on getting all the requirements clearly \\ Always striving to do the best, result in delivering\\ high-quality work  \end{tabular}  &  19/21                                                                                      \\ \cmidrule(l){2-4}   
    
    & (+) Committed team members                                          
    & \begin{tabular}[c]{@{}l@{}} Involve in tasks and complete them beyond expectations \\ Making the other team members motivated to do their\\ work by giving their best \\  Better understanding of requirements from the beginning\\ and manage timelines accordingly \\ Make the developers committed to delivering what \\customers exactly need \end{tabular} & 19/21 \\ \cmidrule(l){2-4} 
        & (+)  Responsible team members                             
        & \begin{tabular}[c]{@{}l@{}}  Full-filling all the assigned tasks result in completing all the tasks \\in one story within a particular sprint \\  Ability to backtrack any issues that occurred in the\\ development of each story\\  Not disregarding each and every task helpful in not\\ missing any requirements \\ Ensure  the delivery of the project within the estimated\\ time and budget \\ Learn required skills and techniques to complete the assigned\\ tasks (e.g., interviewing clients) \end{tabular}  & 18/21                                              \\ \cmidrule(l){2-4} 
    & \begin{tabular}[c]{@{}l@{}} (+) Equality of the team  \end{tabular}                                      & \begin{tabular}[c]{@{}l@{}}   Everyone's contribution in the team is helpful in understanding \\requirements easily \\ Ability to approach to the people easily makes fewer conflicts on tasks\\ Grow the responsibility and commitment towards the project with the\\ sense of ownership resulting in high performance of the tasks \\ Easy to work on dependencies of the functions increasing the\\ progress of the project\end{tabular} & 17/21                                                                                                      \\ \cmidrule(l){2-4} 
    
             & (+) Strong team bond                              
             & \begin{tabular}[c]{@{}l@{}}  Helpful in solving complicated/ massive tasks with less time period \\ Enhance the knowledge sharing among team members \\ Creates positive environment among the team, resulting in a \\high team performance \\  Helpful in having informal discussions to resolve problems \\and make the team learn from each others\end{tabular}      & 17/21     \\\cmidrule(l){2-4} 
        & (+) Transparency of the team                                    
        & \begin{tabular}[c]{@{}l@{}} Avoid potential risks of wrong interpretations of the requirements \\  Get detailed information on changing requirements and their priorities \\ Keeping everyone on the same page, reducing conflicts and \\incomplete tasks \\ Enhance communication and coordination to complete tasks with  trust \end{tabular}     & 15/21                       \\ \cmidrule(l){2-4} 
                
          & (+) Understanding team                                          & \begin{tabular}[c]{@{}l@{}} Ensure to assign tasks based on the expertise area, making early\\ task completion \\  Helpful in understanding technical difficulties and easy negotiation\\s with customers \\  Not being forced to work on tasks, resulting in quality outcome \end{tabular}  & 7/21                                                
          \\ \cmidrule(l){2-4}
          
         & (+/-) Nature of the team                                               
         & \begin{tabular}[c]{@{}l@{}} Energy of the team is helpful in completing tasks even before the \\estimated time \\ Positive feelings (good vibes) of the team reduce insecurities within\\ the team and show the progress of the project\\ Enthusiasm of the team makes the whole development process fun\\ and enjoyable \\ Lazy team members avoid detailed discussions to gain clarity of \\requirements \\ Unprofessional behaviours of the team make an unhappy environment,\\ resulting in incomplete projects   \end{tabular}                                                                                   & 13/21     \\\cmidrule(l){2-4}                     
         & (+/-) Feedback                                                        & \begin{tabular}[c]{@{}l@{}} Helpful in improving user stories and prioritising product backlog \\ Helpful in developing the project in the right direction, keeping\\ customers satisfied \\ Early preparation of the upcoming tasks in accordance with\\ customer needs \\Getting wrong inputs/feedback (especially from customers) \\creates confusion of the product requirements\\ \end{tabular} & 12/21                                                          \\\cmidrule(l){2-4}
         
        & (-) Team conflicts                                                                    & \begin{tabular}[c]{@{}l@{}}  Delays in product delivery as it takes more time than estimated\\ to complete the assigned tasks\\ Refusal to engage with others impact the final outcome \\  Reduce the performance of the whole team and loose \\project interest  \\  Lead to missing requirements due to poor communication \\of conflicted team members \end{tabular}    & 15/21  \\\cmidrule(l){2-4}

  & (-) Poor team management                                  
         & \begin{tabular}[c]{@{}l@{}}  Takes more time to resolve problems with wrong estimations\\ of the project \\  Unable to complete high-priority tasks and create difficulties\\ on the progress of the project \\ Wasting the resources of the project without having any\\ progress/ gain on the project \end{tabular} & 9/21
\\ \midrule

\end{tabular}%
}
\end{table*}

\begin{table*}[]
\caption{Influence of collaborative \& human motivation factors on performing RE-related activities (Customer/Clients/End user-related)}
\label{Table : impact of human-related motivation factors-customers}
\resizebox{\linewidth}{!}{%
\begin{tabular}{@{}clll@{}}
\toprule
\textbf{Sub-Category}                                                                   & \textbf{\begin{tabular}[c]{@{}l@{}}Motivating/ \\ demotivating Factors\end{tabular}}                                          & \textbf{Impact on RE/SE} & \textbf{Responses} \\ \midrule

\multirow{6}{*}{\begin{tabular}[c]{@{}l@{}} Customers\\ Clients\\ End users\end{tabular}} & \begin{tabular}[c]{@{}l@{}}(+) Knowledgeable customers \end{tabular}                 & \begin{tabular}[c]{@{}l@{}} Knowing what they exactly need, helpful in \\getting clear \& detailed requirements \\ Having the right picture of the mind makes \\the designing of the project easier \\  Helpful when working on massive complex \\projects \\ Technical knowledge of the customers make\\ the communication easier and more engagement \\with the development team\end{tabular}  & 20/21       \\ \cmidrule(l){2-4}                                                       
      & (+) Cooperative customers                                                  
       & \begin{tabular}[c]{@{}l@{}} Provides meaningful responses to development\\ team inquiries to understand complex requirements \\ Reduce various project risks in later stages\\ of the project \\ Not getting last minute requirements changes \\ Make the progress of the project smooth \\without conflicts or disturbances \\ Clients' trust towards the project increases\\ and helpful in negotiations related to technical\\ feasibility\end{tabular}  &  19/21                                                                                      \\ \cmidrule(l){2-4}   
    
             & (+) Customer satisfaction                                                 
       & \begin{tabular}[c]{@{}l@{}}  Increase development team's interest and \\engagement towards the project \\Impacts on the organisation, its \\functions and reputation\\ Increase development team's desire to \\complete all the required features  \end{tabular}  &  17/21                                                                                      \\ \cmidrule(l){2-4}               
             & (+)  Customer availability                                            
       & \begin{tabular}[c]{@{}l@{}} Helpful in clarifying unclear requirements \\or getting detailed requirements \\ Development team gets a sense  of understanding\\ of what customers actually want \\ Ability to complete tasks without any delays\end{tabular}  &  16/21                                                                                      \\ \cmidrule(l){2-4}  
             & (+)  Customer feedback                                            
       & \begin{tabular}[c]{@{}l@{}} Acknowledging the work gives a sense of \\accomplishment to the team, resulting in\\ the continuation of the hard work of the team\\ Helpful in developing quality products in \\future projects \\ Get the understanding of requirements related\\ to their specific domain  \end{tabular}  &  14/21                                                                                      \\ \cmidrule(l){2-4}

         &\begin{tabular}[c]{@{}l@{}} (-)  Difficult to handle \\customers \end{tabular}                                           
       & \begin{tabular}[c]{@{}l@{}} Non-stop inquiries and issues from customers\\ slow down the progress of the project\\  Delays the progress of the project \\ Additional pressure of dealing with customers\\ take away practitioners' focus on the real problem\\ and the requirements  \end{tabular}  &  11/21                                                                                      \\ \cmidrule(l){1-4} 
  
\end{tabular}%
}
\end{table*}

\end{document}